\documentclass[pdflatex,sn-mathphys-num]{sn-jnl}

\usepackage{graphicx}%
\usepackage{multirow}%
\usepackage{amsmath,amssymb,amsfonts}%
\usepackage{amsthm}%
\usepackage{mathrsfs}%
\usepackage[title]{appendix}%
\usepackage{xcolor}%
\usepackage{textcomp}%
\usepackage{manyfoot}%
\usepackage{booktabs}%
\usepackage{algorithm}%
\usepackage{algorithmicx}%
\usepackage{algpseudocode}%
\usepackage{listings}%
\usepackage{mathrsfs}

\usepackage[utf8]{inputenc} 
\usepackage[T1]{fontenc}
\usepackage{fullpage}
\usepackage{array}
\usepackage{bm}
\usepackage{caption} 
\usepackage{fancyvrb} 
\VerbatimFootnotes 
\usepackage{float}
\usepackage{framed}
\usepackage{lineno}
\usepackage{mathtools}
\usepackage{multirow}
\usepackage{natbib}
\usepackage{soul}
\usepackage{subcaption} 
\usepackage{tabularx}

\renewcommand{\u}{\boldsymbol{u}}

\newcommand{\F}{\boldsymbol{F}}
\newcommand{\C}{\boldsymbol{C}}
\newcommand{\Sln}{\boldsymbol{T}}
\newcommand{\NOM}{\boldsymbol{P}}
\newcommand{\CAU}{\boldsymbol{\sigma}}
\newcommand{\PP}{\mathbb{P}}
\newcommand{\LL}{\mathbb{L}}

\DeclareMathOperator*{\argmin}{argmin}
\newcommand{\rt}[1]{\textcolor{black}{#1}} 

\begin{document}

\title{Finite-strain constitutive model for shape memory alloys formulated in the logarithmic strain space}

\author[1,2]{\fnm{Alexej} \sur{Moskovka}}

\author[1,3]{\fnm{Martin} \sur{Hor\'{a}k}}

\author[1]{\fnm{Jan} \sur{Valdman}}

\author[4]{\fnm{Michal} \sur{Knapek}}

\author[4]{\fnm{Milo\v{s}} \sur{Jane\v{c}ek}}

\author[5]{\fnm{Petr} \sur{Sedl\'{a}k}}

\author*[5]{\fnm{Miroslav} \sur{Frost}} \email{mfrost@it.cas.cz}

\affil[1]{\orgname{Institute of Information Theory and Automation, Czech Academy of Sciences}, \orgaddress{\street{Pod Vodárenskou v\v{e}\v{z}í 4}, \city{Prague}, \postcode{18200}, \country{Czech Republic}}}

\affil[2]{\orgdiv{Department of Mathematics}, \orgname{Faculty of Applied Sciences, University of West Bohemia}, \orgaddress{\street{Technick\'{a} 8}, \city{Pilsen}, \postcode{30100}, \country{Czech Republic}}}

\affil[3]{\orgdiv{Department of Mechanics}, \orgname{Faculty of Civil Engineering, Czech Technical University in Prague}, \orgaddress{\street{Th\'{a}kurova 7}, \city{Prague}, \postcode{16629}, \country{Czech Republic}}}

\affil[4]{\orgdiv{Department of Physics of Materials}, \orgname{Faculty of Mathematics and Physics, Charles University}, \orgaddress{\street{Ke Karlovu 5}, \city{Prague}, \postcode{12116}, \country{Czech Republic}}}

\affil*[5]{\orgname{Institute of Thermomechanics, Czech Academy of Sciences}, \orgaddress{\street{Dolejškova 5}, \city{Prague}, \postcode{18200}, \country{Czech Republic}}}


\abstract{This work presents a finite-strain version of an established three-dimensional constitutive model for polycrystalline shape memory alloys (SMA) that is able to account for the large deformations and rotations that SMA components may undergo. The model is constructed by applying the logarithmic strain space approach to the original small-strain model, which was formulated within the Generalized Standard Materials framework and features a refined dissipation (rate) function. Additionally, the free energy function is augmented to be more versatile in capturing the transformation kinetics. The model is implemented into finite element software. To demonstrate the model performance and validate the implementation, material parameters are fitted to the experimental data of two SMA, and two computational simulations of SMA components are conducted. The applied approach is highly flexible from the perspective of the future incorporation of other phenomena, e.g., irreversibility associated with plasticity, into the model.}



\maketitle

\section{Introduction}

The unique mechanical properties of shape memory alloys (SMA) have found applications in numerous engineering devices across different industries, e.g., in medicine \cite{Gierig2025}, logistics \cite{JAN-SUB-17}, 
aerospace engineering \cite{Naghipour2024}, and civil engineering \cite{Abavisani2021}. Many of these applications inherently undergo large rotations and/or moderate or finite strains. Consequently, computational analysis of their mechanical response must be extended beyond the small-strain regime, requiring a finite-strain reformulation of the constitutive model.

The traditional approach to dealing with finite strains in phenomenological elastoplasticity is based on the multiplicative decomposition of the deformation gradient \cite{LEE-69}. 
The most common examples of adopting such an approach for modeling SMA mechanics include \cite{HEL-07b, REE-CHR, EVA-SAC-09, ARG-AUR-10, ARG-AUR-11b, STU-PET-12, DAM-LIA-17}. 
The evolution of the elastic part is often governed by various isotropic hyperelastic potentials (e.g. Saint-Venant-Kirchhoff, Neo-Hookean); the remaining term(s) of the decomposition may vary in interpretation and they are related to, e.g., transformation, reorientation, conventional plasticity, viscoplasticity, or non-dissipative non-linear deformation. Certain models incorporate the thermal component in the decomposition, thereby allowing for the simulation of thermomechanical coupling \cite{WAN-ZAK-17, CHR-REE-09, SIE-WUL-22}.

As an alternative approach, the additive decomposition models employing Eulerian logarithmic (Hencky) strain have been developed, see \cite{MUL-BRU-06, XU-LAG-19, XU-LAG-21, ZHA-BAX-21} for SMA-related work. These rate-form models, based on a hypoelastic constitutive relation, benefit from the specific properties of this strain tensor discovered in \cite{XIA-BRU-97}. The additive decomposition of the strain rate tensor, combined
with logarithmic corotational integration, facilitates a straightforward summation of individual contributions to the Eulerian logarithmic strain.

The present approach follows the framework of Miehe, Apel, and Lambrecht \cite{MIE-LAM-02}, employing an additive decomposition of the logarithmic strain. This contrasts with the models discussed above, which are based on
a decomposition of the logarithmic strain rate tensor. Its modular structure allows a simple extension of constitutive structures from the geometrically linear theory to the finite-strain theory using purely geometric transformations. Namely, the finite-strain model is built by reinterpreting the inputs of the original small-strain model within the Lagrangian logarithmic (Hencky) strain space and employing geometric transformations to its outputs to obtain a chosen (Lagrangian or Eulerian) stress measure and the corresponding stiffness tensor.

The approach has been successfully applied to many constitutive models of metal plasticity \cite{Aldakheel2017} and damage \cite{Horak2013, Holthusen2022}, polymers \cite{Miehe2011}, etc., 
and its limitations have also been addressed \cite{FRI-STE-22}. In the realm of SMA models, the methodology has been only recently applied in \cite{WOO-KAL-24} and \cite{CHA-KAL-24} for extending small-strain models \cite{SOU-98} and \cite{WOO-KAL-22} to finite strains, respectively.

In this work, we benefit from the modular setting and develop the finite-strain version of the small-strain SMA model originally introduced in \cite{SED-FRO-IJOP, FRO-BEN-MMS}. The model was formulated within the Generalized Standard Materials framework \cite{HAL-NGU}; hence, it perfectly fits the approach devised in \cite{MIE-LAM-02}. It features an independent description of phase transformation and (re)orientation (including self-accommodation of martensite), as well as an enhanced dissipation (rate) function that naturally captures the mechanical stabilization of martensite \cite{LIU-FAV, BEL-RES-20}, and tension-compression asymmetry \cite{LIU-XIE}. Later versions focused on intermediate phase transformations in NiTi \cite{MD-NITIFE}, localization of transformation \cite{FRO-BEN-IJSS}, or refined description of elastic properties \cite{JIMSS-SPRING}. The predictive capabilities of the model were validated against mesoscale experimental data in \cite{SMS-SNAKE, SMS-DSCT, IJSS-DSCT}. Here, the original model is adopted in the logarithmic strain framework, with particular attention to its ability to cope with large rotations and moderate strains and with the perspective of using the framework to comply with large rotations and finite strains in concurrently developed constitutive model tailored for complex plastic deformation processes in NiTi motivated by \cite{HEL-SED-19}.

Using the implementation to the finite element method, some illustrative computational simulations of NiTi structures are performed. To assess the robustness of the final model with respect to its parametrization, we also conduct the simulations with a different set of material parameters, whose particular choice is motivated by our research on another family of superelastic/shape memory alloys. Metastable $\beta$-Ti alloys based on Ti-Zr-Nb-Sn exhibit promising superelastic properties \cite{HOS-MIY-15}, which are utilizable in medical applications. Moreover, they are free of nickel, which is deemed potentially hazardous to health \cite{
Koester2000}. In this respect, we have modified the energetic function of the core model to get a more flexible description of transformation kinetics.

\section{SMA Model Description}
\label{sec:model}

The core of the small-strain SMA model introduced in \cite{SED-FRO-IJOP} is converted into a finite-strain model using the logarithmic strain space approach introduced in \cite{MIE-LAM-02}. The advantage of this approach is that small-strain models can be seamlessly integrated into the finite-strain formulation without requiring substantial modifications. It consists of three steps. In the first step, the information on the deformation of the body is used to compute the logarithmic strain measure. In the second step, this strain measure enters a constitutive law, which may retain the identical structure as in the small-strain formulation. In the third step, the generalized stress tensor and the algorithmic stiffness tensor (consistent tangent modulus) provided by the model are transformed into a suitable stress measure and the corresponding stiffness tensor, respectively. The transformation is essentially of a geometric nature and employs specific projection tensors. The other computational procedures, e.g., handling internal variables or resolving the constitutive response (time discretization scheme, state update procedure), can be retained from the original small-strain model.

\subsection{Preliminaries}
\label{subsec:model-pre}

Given a displacement mapping $\u$, we assume the corresponding deformation gradient $\F = \nabla \u + I$ and the (symmetric) right Cauchy-Green deformation tensor
\begin{equation}
\C = \F^T \F.
\end{equation}
Following the ideas from \cite{MIE-LAM-01, MIE-LAM-02}, we work with the logarithmic strain measure, known as the Lagrangian Hencky strain (also called the logarithmic strain) defined through
\begin{equation} \label{eq:log-str}
    \bm{H} = \frac{1}{2} \ln(\C).
\end{equation}

Consider now the multiplicative decomposition of the deformation gradient to an elastic part, $\bm{F}^{\rm el}$, and inelastic part, $\bm{F}^{\rm in}$, as $\bm{F} = \bm{F}^{\rm el}\bm{F}^{\rm in}$. Under the assumption of co-axiality of $\bm{F}^{\rm in}$ and $\bm{F} = \bm{F}^{\rm el}\bm{F}^{\rm in}$, one gets
\begin{equation} \label{eq:add-ansatz}
     \bm{H} = \frac{1}{2} \ln ({\bm{F}^{\rm in}}^{\top}{\bm{F}^{\rm el}}^{\top}\bm{F}^{\rm el}\bm{F}^{\rm in}) = \bm{H}^{\rm el} + \bm{H}^{\rm in}.
\end{equation}
To arrive at a computationally attractive framework, the logarithmic strain approach assumes the validity of this relation beyond the co-axility premise \cite{MIE-LAM-02}. As demonstrated in previous works, e.g., \cite{MIE-LAM-02, CHA-KAL-24}, the results obtained with the additive ansatz \eqref{eq:add-ansatz} are ``very close" to those obtained with the corresponding models based on multiplicative decomposition, see also \cite{FRI-STE-22}.

\subsection{Formulation within the Logarithmic Strain Space}
\label{subsec:model-log-str}

The finite-strain model is constructed via transforming the existing small-strain SMA model introduced in \cite{SED-FRO-IJOP, FRO-BEN-MMS} through the logarithmic strain space approach. Here, we briefly describe the main features of its constitutive core. The reader is referred to \cite{SED-FRO-IJOP, FRO-BEN-MMS} for more details on the physical motivation and numerical implementation, to \cite{BEN-FRO-DCDSS, FRO-BEN-IJSS, MD-NITIFE} for extensions of the small-strain model, and to \cite{JMEP-SPRING, JMEP-STENT, SMS-SNAKE, SMS-DSCT} for examples of model performance and validation.

The microstructural state of martensite is assumed to be characterized by two internal variables: the volume fraction of martensite (scalar), $\xi^{\rm M}$, and transformation strain tensor of martensite, $\bm{H}^{\rm M}$. We require that $\bm{H}^{\rm M}$ is symmetric,  i.e., $\bm{H}^{\rm M} \in \mathbb{R}^{3 \times 3}, H^{\rm M}_{ij} = H^{\rm M}_{ji}, \, i,j \in \{1,2,3\}$, and traceless, i.e., ${\rm tr}(\bm{H}^{\rm M}):= \sum^3_{i=1} H^{\rm M}_{ii} = 0$, to assure the volume-preserving property of the transformation, cf. \cite{Holthusen2022}. The additive decomposition of the total strain, $\bm{H}$, to an elastic part, $\bm{H}^{\rm el}$, and a transformation-related (inelastic) part, $\bm{H}^{\rm tr}$, is expected as follows:
\begin{equation}\label{eq:str-decomp}
\bm{H} = \bm{H}^{\rm el} + \bm{H}^{\rm tr} = \bm{H}^{\rm el} +\xi^{\rm M} \bm{H}^{\rm M}.
\end{equation}

The natural constraint on the volume fraction of martensite is complemented by directional and loading mode restrictions on transformation strain,
\begin{equation}\label{eq:constraints}
0 \leq \xi^{\rm M} \leq 1, \qquad \langle \bm{H}^{\rm M} \rangle \leq k,
\end{equation}
where $k$ is a material parameter representing the maximum transformation strain of martensite attainable in uniaxial tensile loading and the material function $\langle \cdot \rangle$ takes a specific form proposed in \cite{SED-FRO-IJOP}:
\begin{equation}
\label{eq:convex-set}
\langle \bm{H}^{\rm M} \rangle = I_2(\bm{H}^{\rm M}) \frac{\cos\left(\frac{1}{3}{\arccos(1-a(I_3(\bm{H}^{\rm M}) + 1))}\right)}{\cos\left(\frac{1}{3}{\arccos(1-2a)}\right)},
\end{equation}
where
\begin{equation}
I_2(\bm{H}^{\rm M}) = \sqrt{\frac{2}{3}\bm{H}^{\rm M}:\bm{H}^{\rm M}}, \quad I_3(\bm{H}^{\rm M}) = 4\frac{\det(\bm{H}^{\rm M})}{I_2(\bm{H}^{\rm M})^3}.
\end{equation}
The material parameter $0 \leq a \leq 1$ accounts for the tension-compression asymmetry.

The core of the model consists of two particular functions. The energy function (thermodynamic potential), $f^T$, consisting of elastic, chemical, and internal contributions, takes the form
\begin{eqnarray}\label{eq:free-en}
f^T(\bm{H},\bm{H}^{\rm M},\xi^{\rm M})
 & = & \frac{1}{2}K{\rm tr}(\bm{H})^2 + G(\xi^{\rm M})\|{\rm dev}({\bm{H}}) - \xi^{\rm M} \bm{H}^{\rm M}\|^2 \nonumber \\
 &   & +\; \Delta s^{AM}(T-T_0)\xi^{\rm M} \nonumber \\
 &   & +\; r^{\rm int}(\xi^{\rm M}, \bm{H})
\end{eqnarray}
where the bulk modulus, $K$, is assumed phase-independent, the (apparent) shear modulus of the phase mixture is given by the Reuss rule
\begin{equation}
    G(\xi^{\rm M}) = \frac{G^{\rm A}G^{\rm M}}{\xi^{\rm M} G^{\rm A} + (1-\xi^{\rm M}) G^{\rm M}},
\end{equation}
$\Delta s^{AM}$ denotes the specific entropy change associated with the phase transformation, $T_0$ is the corresponding phase-equilibrium temperature (see \cite{SED-FRO-IJOP} for details), and $\|\cdot\|$ denotes the Frobenius norm. Temperature, $T$, is considered a prescribed (albeit variable) parameter, as the superscript indicates.

The last term in \eqref{eq:free-en} represents phenomenological energy contributions associated with the heterogeneity of the material caused by its multiphase state and/or by its polycrystalline nature,
\begin{equation}\label{eq:en-int}
r^{\rm int} = \frac{1}{2}E^{\rm hard}\xi^{\rm M} \langle \bm{H}^{\rm M} \rangle^{2} + E^{\rm kin}_0(1-\xi^{\rm M})^{n_0} + E^{\rm kin}_1(\xi^{\rm M})^{n_1},
\end{equation}
where $E^{\rm hard}, E^{\rm kin}_0, E^{\rm kin}_1$ and $1 < n_0, n_1 \leq 2$ are material parameters obtainable from experimental data. The form of the first term on the right-hand side is motivated by \cite{CHE-DUV}, where such a type of internal energy reflecting hardening during martensite reorientation was derived from micromechanical considerations. The other terms are inspired by the work \cite{LAG-CHE}, where they were associated with multiphase composition (“energy of mixing,” cf. \cite{PAT-LAG-II}); they enable to capture the phase transformation kinetics better.

The dissipation function, $d$, in which both transformation and reorientation of martensite are considered, takes the form
\begin{eqnarray}\label{eq:dissip}
\dot\xi^{\rm M} \geq 0 \quad \Rightarrow \quad d(\bm{H},\xi^{\rm M},\dot{\bm{H}}^{\rm M},\dot{\xi}^{\rm M})
& = &\;\Delta s^{AM}[(T_0 - M_{\rm s}) + \xi^{\rm M} (M_{\rm s} - M_{\rm f})]\dot{\xi}^{\rm M} \nonumber \\
& & +\;\sigma^{\rm reo}\|\dot{\xi}^{\rm M}{\bm H}^{\rm M} + \xi^{\rm M}\dot{\bm H}^{\rm M}\|, \\
\dot\xi^{\rm M} < 0 \quad \Rightarrow \quad d({\bm H},\xi^{\rm M},\dot{\bm H}^{\rm M},\dot{\xi}^{\rm M})
& = & \Delta s^{AM}[(T_0 - A_{\rm f}) + \xi^{\rm M} (A_{\rm s} - A_{\rm f})]\dot{\xi}^{\rm M} \nonumber \\
& & +\;\sigma^{\rm reo}\left[|\dot{\xi}^{\rm M}|\cdot \|{\bm H}^{\rm M}\| +
\xi^{\rm M}\cdot \|\dot{\bm H}^{\rm M}\| \right], \label{eq:dissip-2}
\end{eqnarray}
with $M_{\rm s},M_{\rm f},A_{\rm s}$ and $A_{\rm f}$ being material constants related to stress-free transformation temperatures and $\sigma^{\rm reo}$ is related to the critical stress for martensite reorientation. Let us note that the above rate-independent dissipation function allows capturing (re)orientation processes and the mechanical stabilization of martensite, cf. \cite{FRO-VAL-22}; hence, it differs for forward \eqref{eq:dissip} and reverse \eqref{eq:dissip-2} transformation.

After establishing the internal variables in \eqref{eq:constraints} and defining the two constitutive functions in \eqref{eq:free-en}, \eqref{eq:dissip} and \eqref{eq:dissip-2}, the model formulation can be completed using the framework of Generalized Standard Materials/Media (GSM) \cite{HAL-NGU}, which establishes the relation between thermodynamic forces and fluxes via a generalized Biot's evolutionary equation. For the (convex) model at hand, it breaks down to the relation for the logarithmic stress
\begin{equation}\label{eq:stress}
 \bm{T} = \frac{\partial f(\bm{H},\bm{\alpha}^{\rm M})}{\partial {\bm H}}, 
\end{equation}
and the governing principle for the evolution of internal variables  
\begin{eqnarray}\label{eq:minimiz}
0 \in \partial_{\bm{\alpha}^{\rm M}} f^T(\bm{H},\bm{\alpha}^{\rm M}) + \partial_{\dot{\bm{\alpha}}^{\rm M}}\, d^{\rm T}(\bm{\alpha}^{\rm M},\dot{\bm{\alpha}}^{\rm M}), \quad \bm{\alpha}^{\rm M}(0) = \bm{\alpha}^{\rm M}_0,
\end{eqnarray}
with the vector of internal variables $\bm{\alpha}^{\rm M} := (\bm{H}^{\rm M},\xi^{\rm M})$ initially taking the value $\bm{\alpha_0}^{\rm M}$. Symbols $\partial$ and $\in$ denote the subdifferential and inclusion relation, respectively, see \cite{MIE-LAM-02}. Let us note that GSM framework ensures -- under some simple conditions satisfied by \eqref{eq:dissip} and \eqref{eq:dissip-2} -- the thermodynamic consistency of the proposed constitutive model \cite{HAL-NGU, MIE-LAM-02}, and that establishing the constitutive law via \eqref{eq:minimiz} provides an alternative to the defining the domain(s) of admissible force(s) together with flow rule(s) as common for rate-independent models \cite{HAC-FIS}.

\subsection{Constitutive State Update and Post-processing of Outputs}
\label{subsec:model-post}

The time-continuous version of the model introduced in the previous section is discretized in time via backward Euler scheme ($\dot\xi^{\rm M} \rightarrow (\xi^{\rm M} - \xi^{\rm M}_{k-1})/\Delta t, \, \dot{\bm{H}}^{\rm M} \rightarrow (\bm{H}^{\rm M} - \bm{H}^{\rm M}_{k-1})/\Delta t$ with time increment $\Delta t$), see \cite{FRO-BEN-MMS} for details, hence the time-discrete form of the constitutive law at a time instant $t_k$ takes the form of a constrained (convex) minimization problem
\begin{eqnarray}\label{TIP}
\bm{\alpha_k} = \argmin_{\bm{\alpha}} \{f^{T_k}(\bm{\alpha}) + d_k(\bm{\alpha},\bm{\alpha}_{k-1}) + \mathcal{I}(\bm{\alpha},\bm{\alpha}_{k-1})\}. 
\end{eqnarray}
The subscript $k$ denotes the discretized contributions corresponding to the discrete variational formulation specified in \cite{MIE-LAM-02}, $\mathcal{I}$ is the indicator function, which ensures satisfaction of the constraints \eqref{eq:constraints} \cite{SED-FRO-IJOP}, see \cite{FRO-VAL-22} for numerical implementation details. 

After resolving the problem \eqref{TIP}, the logarithmic stress tensor is explicitly computed through (the discretized version of) equation \eqref{eq:stress}. The constitutive update needed for finite element implementation is subsequently completed by computing the consistent tangent modulus, i.e. the discretized version of the relation ${\rm d}\bm{T} / {\rm d}\bm{H}$; the analytical form can be derived in the very same manner as in \cite{SED-FRO-IJOP}, see also \cite{STU-PET-12}.

These two model outputs are consequently transformed into the non-symmetric nominal stress, $\bm{P}$, and nominal stiffness tensor (tangent modulus), $\mathbb{M}$, respectively, which directly enter the finite element formulation of the incremental boundary-value problems. The algebraic relations
\begin{equation}\label{eq:nom-str-stiff}
    \NOM = \Sln : \PP \, , \qquad \mathbb{M} = \PP^T : \frac{{\rm d} \Sln}{{\rm d}\bm{H}} : \PP \, + \, \Sln : \LL \, ,
\end{equation}
use the fourth- and sixth-order transformation (projection) tensors
\begin{equation} \label{eq:proj-tens}
    \PP = \frac{\partial \bm{H}}{\partial \F}, \qquad \LL = \frac{\partial^2 \bm{H}}{\partial \F \partial \F}.
\end{equation}
The assembly of the transformation tensors $\PP$ and $\LL$ from \eqref{eq:proj-tens} is more technical and is detailed in \cite{MIE-LAM-02}. The nominal stress $\NOM$ can be used to calculate 
the Cauchy stress $\CAU$ given by
\begin{equation}\label{eq:stresses}
\bm{\sigma} = \frac{1}{\det(\F)} \, \NOM \F^T.
\end{equation}

\section{Finite Element Implementation and Computational Examples}
\label{sec:simul}

As shown in \cite{MIE-LAM-02}, the variational formulation of the constitutive law can be easily combined with the principle of minimum potential energy to yield a global variational formulation for the incremental boundary-value problem of a material body. The resulting ``combined" minimization problem (cf. \cite{HAC-FIS}) can be solved in a monolithic manner, e.g., \cite{FRO-VAL-22}, or via the alternating minimization approach, e.g., \cite{FRO-BEN-MMS}, which fits well into common implementations of finite element methods. The latter way has also been used in this work, as described below. The material parameters of the constitutive model are then fitted to particular alloys and tested on benchmark computational examples.

\rt{\subsection{Finite Element Implementation}}

\rt{The logarithmic strain model of SMA has been implemented into the finite element method in the MATLAB computational environment. The implementation builds on a lightweight open-source finite element code developed in \cite{CSV-19}. The original code: i) features the innovative assembly of FEM matrices (discretized versions of operators), their memory-saving storage and efficient handling via full vectorization, ii) has a modular structure, which allows to alter the employed constitutive routine (state update procedure) in the same fashion as user material routines are provided in common commercial finite element packages (e.g. UMAT in Abaqus, USERMAT in Ansys, HYPELA2 in MSC MARC), and iii) is continuously extended, improved, and streamlined, cf. recent papers \cite{Pospisil2023, Svetlik2024, Karatson2025}.}

\rt{The fundamental structure of the code -- boundary value problems are resolved via the finite element method with implicit time discretization and (semismooth) Newton solution algorithm for the linearized system -- was retained. However, the following major modifications had to be made with respect to the finite-strain SMA model: 1) the elastoplastic constitutive routine (state update procedure) was substituted by the finite-strain SMA model described in Subsection \ref{subsec:model-log-str}, 2) the pre- and post-processing phases introduced in Subsections \ref{subsec:model-pre} and \ref{subsec:model-post} were added to the main routine of the code; this required the substitution of the strain-displacement operator (matrix) by the deformation gradient-displacement operator and the construction of the projection tensors expressed by Eqs. \eqref{eq:nom-str-stiff} and \eqref{eq:proj-tens}, 3) the robustness of the code was increased by implementing an adaptive time-incrementation procedure, 4) the flexibility of the code was enhanced by the option to prescribe temperature within the integration (Gauss) points; this is particularly beneficial in SMA (isothermal) simulations, where the temperature is an additional, user-controlled parameter that fundamentally affects the mechanical response.}

\rt{The main modification from the point of view of SMA modeling is the complete substitution of the constitutive routine listed above as 1). In their original work \cite{CSV-19}, the authors formulated two elastoplastic constitutive models using the conventional flow plasticity theory. Their state update procedure relied on the elastic predictor–plastic corrector (return mapping) solution algorithm. The proposed SMA model builds on works \cite{SED-FRO-IJOP, FRO-BEN-MMS} using GSM and its time-discrete formulation takes the compact form of the minimization problem presented in Eq. \eqref{TIP}. Although this problem can be transformed into a system solvable with return mapping techniques, the minimization formulation naturally leads to a state update procedure based on (non-smooth) mathematical optimization. In the present study, the Matlab in-built function} \verb"fminsearch" \rt{based on the Nelder-Mead minimization algorithm is employed; this algorithm appeared robust and reliable in our previous studies. In future work, we plan to assess the performance of other minimization methods, e.g., bundle or non-smooth Newton methods. A Matlab-coded toolbox for automatic differentiation of mathematical functions called ADiGator was used to compute the consistent tangent modulus \cite{ADiGator}. The robustness of the Newton method was reinforced by applying some regularization to functions $\langle \cdot \rangle$ and $d$ (defined in Eqs. \eqref{eq:convex-set} and \eqref{eq:dissip}--\eqref{eq:dissip-2}, respectively). All numerical aspects of modifications of the original code stemming from its finite-strain extension will be detailed in a forthcoming paper \cite{Moskovka2025}. \medskip}

\subsection{Material Parameters}

To obtain realistic quantitative results in the next subsections, the material parameters of the model have been fitted to experimental data from two polycrystalline alloys with distinct transformation characteristics, such as kinetics and attainable transformation strain. The first is a conventional cold-drawn NiTi wire (FWM NiTi\#1–CW \cite{FWM}), which is widely used in applications; the other one is a Ni-free metastable titanium alloy with the nominal composition Ti-18Zr-11Nb-3Sn (at.\%); it is a member of a family of quaternary alloys considered a prospective alternative to NiTi in medical applications \cite{HOS-MIY-15, JIA-GLO-24}.

\begin{figure}[h]
\centering
\includegraphics[width=0.95\textwidth]{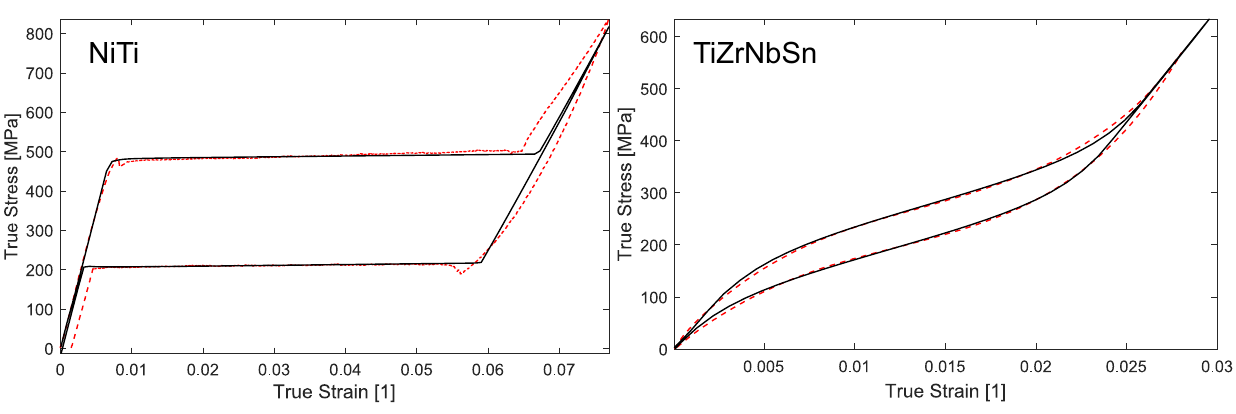}
\caption{Experimental data on uniaxial tension (red dashed line) of NiTi at $40\,^{\circ}{\rm C}$ (left) and Ti-18Zr-11Nb-3Sn at $23\,^{\circ}{\rm C}$ (right) alloys compared to their simulation counterparts (black solid line).}\label{fig:mater_stress-strain}
\end{figure}

The superelastic stress-strain relation for NiTi at $40\,^{\circ}{\rm C}$ depicted in red dashed line in the left panel of Fig.~\ref{fig:mater_stress-strain} was adopted from \cite{SMS-SNAKE}, the same relation for Ti-18Zr-11Nb-3Sn alloy at $23\,^{\circ}{\rm C}$ can be seen in the right panel of Fig.~\ref{fig:mater_stress-strain}. The material parameters of NiTi alloys were fitted based on the hysteresis loop in Fig.~\ref{fig:mater_stress-strain} and further experimental characterization performed in \cite{SMS-SNAKE}, see Table~\ref{tab:params}.\footnote{Let us note that the physical origin of the flat stress plateau observed for NiTi alloy in experiments -- localization of the transformation (``shearband'') \cite{SCIENCE} -- is usually disregarded in general SMA modeling and the material parameters are fitted as the plateau would be the true material response with very weak hardening. Since the choice of the two experimental characteristics in this work is motivated by the intention to demonstrate the capability to capture very different transformation kinetics, we will stick to this (over)simplifying approximation. A rigorous treatment of this issue was proposed, e.g., in our previous work \cite{FRO-BEN-IJSS}.} The black solid line shows the corresponding mechanical response as simulated by the model.  

The metastable Ti-18Zr-11Nb-3Sn alloy was manufactured in the same manner as in \cite{KIM-MIY-18}. The tensile response was obtained on a dog-bone cylindrical sample (3 mm in diameter) with Instron 5882 deformation machine equipped with Epsilon 3442 LHT extensometer at quasistatic deformation rate of $5 \cdot 10^{-4}$ s$^{-1}$ after about 25 cycles of a stabilization (tensile) training up to $0.04$ strain, cf. \cite{KON-WAN-22}. As a more complex experimental characterization of Ti-18Zr-11Nb-3Sn alloy is currently underway, the experimental measurements in Fig.~\ref{fig:mater_stress-strain} were complemented by data obtained from the alloy with the same composition and manufacturing/processing conditions in \cite{KON-WAN-22}. Values of parameters in Table \ref{tab:params} with missing experimental information were estimated based on the properties of a similar alloy -- the bulk modulus was estimated based on \cite{KIM-MIY-18} -- or were conservatively adjusted -- the tension-compression asymmetry factor was set to zero. Again, the simulated response is marked with a black solid line in Fig.~\ref{fig:mater_stress-strain}.

\begin{table}[]
\caption{\label{tab:params}Table of material parameters used in finite-element computations. The number in the last column refers to the relation(s) in which the parameter appears.}
\begin{center} 
\renewcommand{\arraystretch}{1.05}
{\begin{tabular}{@{}lllll}
& & & \\ 
\hline
Parameter & Unit & NiTi & Ti-18Zr-11Nb-3Sn & Equation no.\\
\hline \\[-4.0mm]
$k$ & {[}1{]} & $0.06$ & $0.018$ & \eqref{eq:constraints} \\
$a$ & {[}1{]} & $0.97$ & $0$ & \eqref{eq:convex-set} \\
$K$ & {[}GPa{]} & $148$ & $120$ & \eqref{eq:free-en} \\
$G^{\rm A},G^{\rm M}$ & {[}GPa{]} & $25, 12$ & $14, 15.5$ & \eqref{eq:free-en} \\
$\Delta s^{\rm AM}$ & {[}MPa/$^\circ$C{]} & $0.34$ & $0.041$ & \eqref{eq:free-en}, \eqref{eq:dissip}, \eqref{eq:dissip-2}\\
$T_0$ & {[$^\circ$C]} & $-17$ & $-61$ & \eqref{eq:free-en}, \eqref{eq:dissip}, \eqref{eq:dissip-2} \\
$E^{\rm hard}_{\rm MA}$ & {[}MPa{]} & $1.79$ & $0.18$ & \eqref{eq:en-int}\\
$E^{\rm kin}_0, E^{\rm kin}_1$ & {[}MPa{]} & $0, 0$ & $5.7, 5.7$ & \eqref{eq:en-int}\\
$n_0, n_1$ & {[}1{]} & $0, 0$ & $1.1, 1.1$ & \eqref{eq:en-int}\\
$M_{\rm s},M_{\rm f}$ & {[$^\circ$C]} & $-23, -25$ & $-61, -76$ & \eqref{eq:dissip}\\
$A_{\rm s},A_{\rm f}$ & {[$^\circ$C]} & $-13, -10$ & $-61, -46$ & \eqref{eq:dissip-2} \\
$\sigma^{\rm reo}$ & {[}MPa{]} & $100$ & $20$ & \eqref{eq:dissip}, \eqref{eq:dissip-2} \\
\hline
\end{tabular}}
\end{center}
\end{table}

\subsection{Simulation 1: Bending of a Cantilever Beam}

In the first example, we simulate the bending of a slender cantilever beam in an initially horizontal position. The beam has a square cross-section ($0.5 \times 0.5$ mm$^2$) and a length-to-height ratio of 30. It is divided into a uniform mesh of 3,840 ($8 \times 8 \times 60$) linear hexahedral (Q1) 
elements. The boundary conditions are such that the nodes at one end of the beam are completely fixed while the other end of the beam is loaded in the vertical direction to a maximum displacement of 1 cm, giving rise to a bending moment. No prescribed volume forces are assumed; the temperature of the simulations is the same as for the superelastic curves in Fig.~\ref{fig:mater_stress-strain}.

\begin{figure}[h]
\centering
\includegraphics[width=0.95\textwidth]{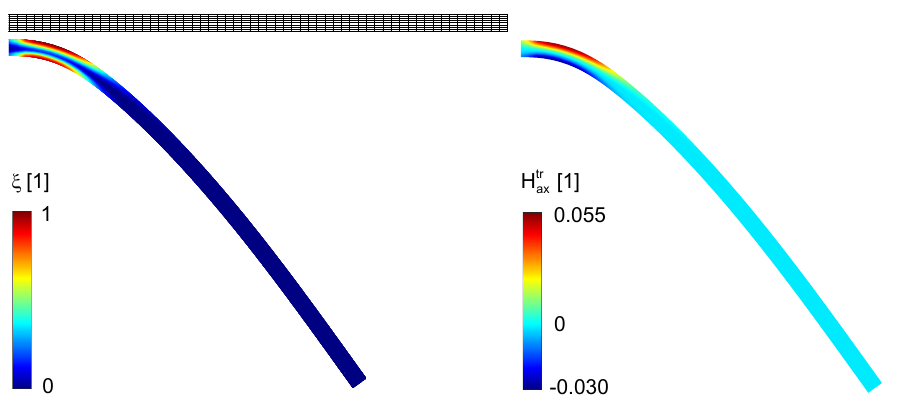}
\caption{Finite-element simulation of deflection of a NiTi cantilever beam. Distribution of volume fraction of martensite (left) and the longitudinal component of transformation strain tensor (right).}\label{fig:cantilever_NiTisnk_distrib}
\end{figure}

Figure \ref{fig:cantilever_NiTisnk_distrib} shows the spatial distribution of the volume fraction of martensite, $\xi$, and the component of the transformation strain in the direction of the longitudinal axis, $\bm{H}^{\rm tr}_{\rm ax}$, in the most deformed state for the NiTi beam. The initial configuration with the overlaid mesh is also plotted. The mechanical response is dominated by bending. The transformation process initiates near the fixed end and spreads towards the free end. The maximum absolute value of the transformation strain component reached in tension on the upper surface differs substantially from that reached in compression on the lower surface. This is a direct consequence of the asymmetry parameter, $a$ in \eqref{eq:convex-set}, which also leads to the asymmetry in the distribution of martensite through the beam cross-section in the transforming region.

\begin{figure}[h]
\centering
\includegraphics[width=0.95\textwidth]{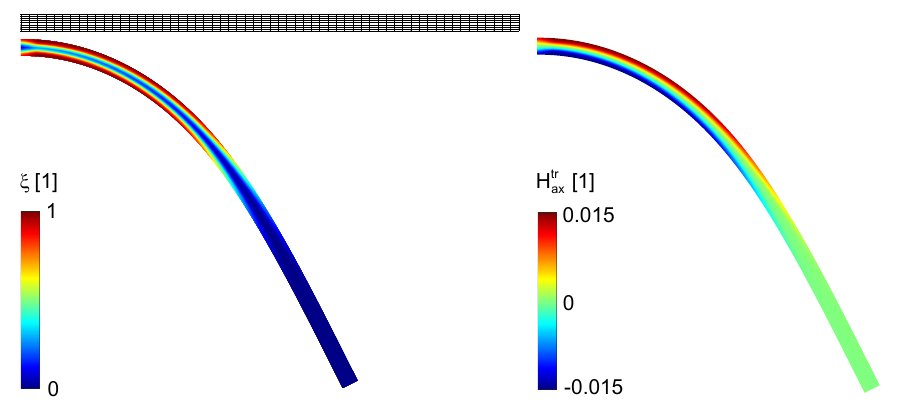}
\caption{Finite-element simulation of deflection of a Ti-18Zr-11Nb-3Sn cantilever beam. Distribution of volume fraction of martensite (left) and the longitudinal component of transformation strain tensor (right).}\label{fig:cantilever_TiNbZrSn_distrib}
\end{figure}

\begin{figure}[h]
\centering
\includegraphics[width=0.95\textwidth]{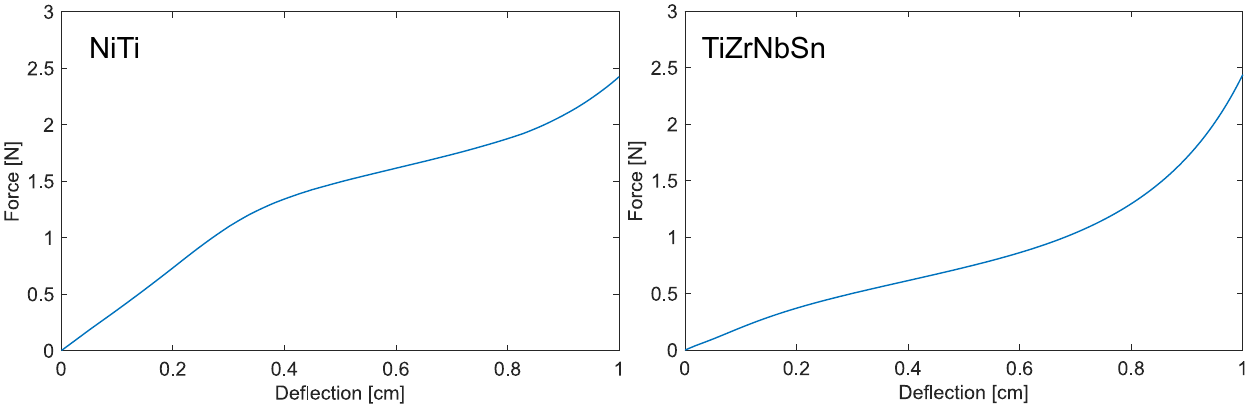}
\caption{The evolution of force with deflection from finite-element simulations of NiTi (left) and Ti-18Zr-11Nb-3Sn (right) cantilevers.}\label{fig:cantilevers_F-u}
\end{figure}

The simulation counterpart for Ti-18Zr-11Nb-3Sn alloy is presented in Fig.~\ref{fig:cantilever_TiNbZrSn_distrib}. Compared to the first simulation, the phase-transformed region propagates further along the beam, but its curvature is lower there, and spatial distributions are more symmetric. This can be rationalized by the significantly lower value of the parameter $k$ (related to the maximum attainable transformation strain in uniaxial tension) and the zero asymmetry parameter, $a$. Finally, Fig.~\ref{fig:cantilevers_F-u} shows the evolution of the (vertical component of) force at the free end with displacement for both simulations. Although the deflection at the end of loading is identical for both cases, the curves differ since the respective alloys exhibit different transformation kinetics (cf. Fig.~\ref{fig:mater_stress-strain}).

\subsection{Simulation 2: Stretching of a Helical Spring Coil}

Computational analysis of the deformation of a helical spring is a common benchmark of finite-strain SMA models
\cite{ARG-AUR-11b, STU-PET-12, DAM-LIA-17, ZHA-BAX-21}. 
We simulate the stretching of a single coil of a helical spring at a constant temperature (corresponding to the superelastic curve in Fig.~\ref{fig:mater_stress-strain}), imposing periodic boundary conditions with a shifted axial displacement to ensure that the geometry represents one segment of an infinitely long virtual spring. The coil is wound from a wire with a circular cross-section of 0.1 mm in diameter. The outer diameter of the coil is 2.1 mm (i.e., the spring index 20), and the initial pitch is 0.1 mm (i.e., the initial pitch angle 3.7$\,^\circ$). The finite element model comprises 4,320 linear hexahedral elements with 48 elements per cross-section distributed uniformly in 90 layers along the wire axis. Stretching is simulated with a gradual change of prescribed displacement in the coil axis-direction
of the central node at one end, while displacements
in the other directions are fixed. The central node at the other end is fixed completely. Periodic boundary conditions with a shifted axial displacement are imposed at the remaining nodes of both ends (cross-sections) of the coil. The rigid body motion of the whole coil is prevented. Hence, the simulated response is the same in any cross-section as required by the infinite spring assumption.

\begin{figure}[h]
\centering
\includegraphics[width=0.95\textwidth]{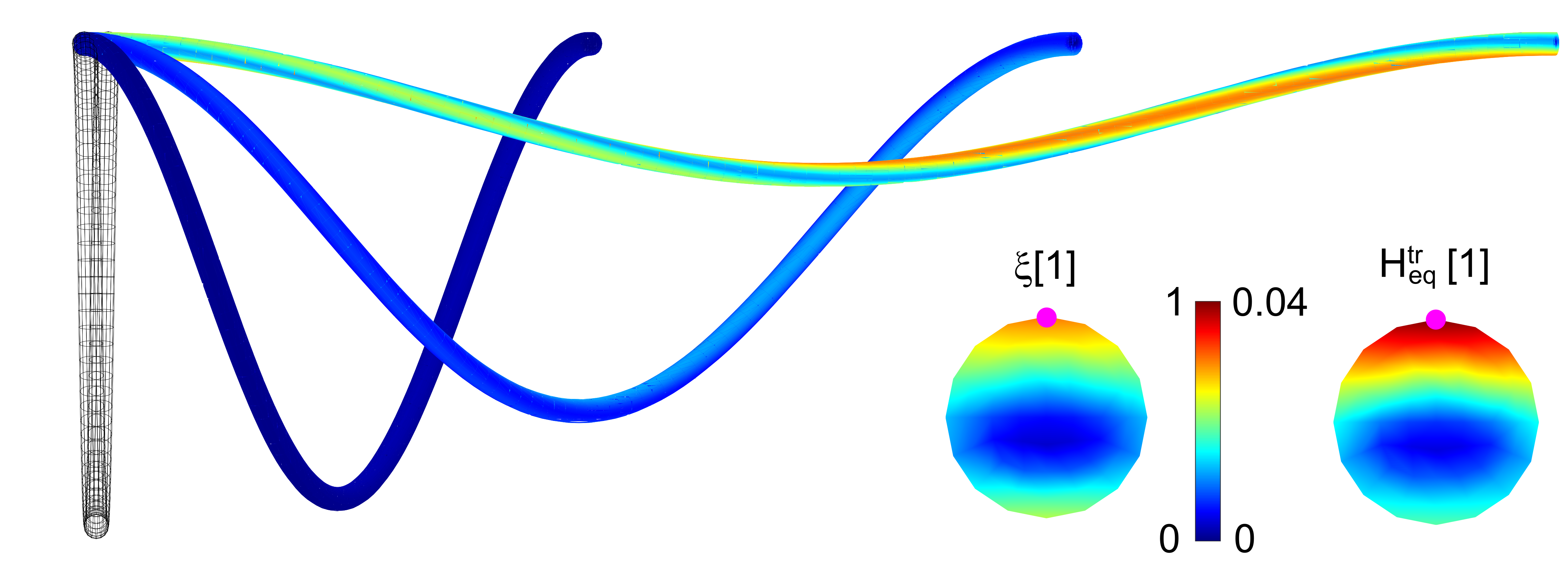}
\caption{Finite-element simulation of a NiTi coil stretching. Initial and three deformed configurations -- 2, 4 and 6 mm strokes -- with the distribution of the volume fraction of martensite on the surface. The inset shows the distribution of martensite, $\xi$, and equivalent transformation strain, $\bm{H}^{\rm tr}_{\rm eq}$, within an arbitrary cross-section at 6 mm stroke. The surface point marked by a magenta circle is the closest to the coil axis.}\label{fig:coil_NiTisnk_distrib}
\end{figure}

\begin{figure}[h]
\centering
\includegraphics[width=0.95\textwidth]{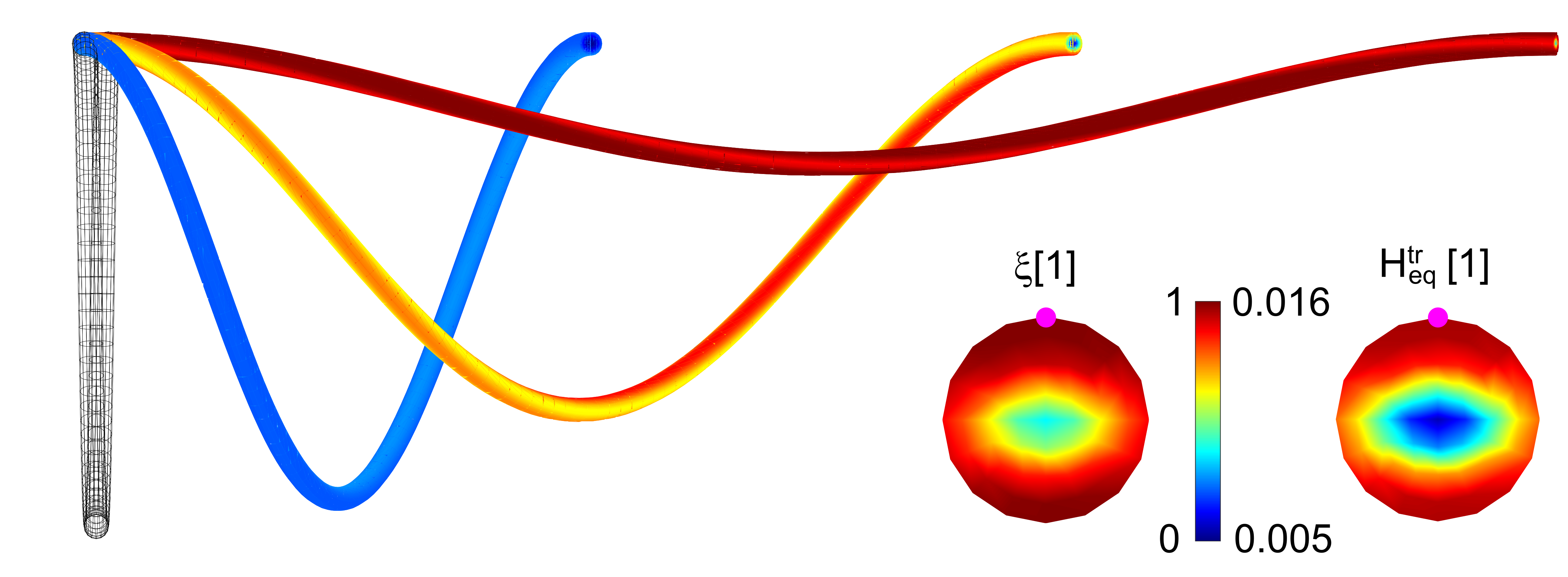}
\caption{Finite-element simulation of a Ti-18Zr-11Nb-3Sn coil stretching. Initial and three deformed configurations -- 2, 4 and 6 mm strokes -- with the distribution of volume fraction of martensite on the surface. The inset shows the distribution of martensite, $\xi$, and equivalent transformation strain, $\bm{H}^{\rm tr}_{\rm eq}$, within an arbitrary cross-section at 6 mm stroke. The surface point marked by a magenta circle is the closest to the coil axis.}\label{fig:coil_TiNbZrSn_distrib}
\end{figure}

Figure~\ref{fig:coil_NiTisnk_distrib} shows the initial position and three deformed configurations with increasing stroke (2, 4, and 6 mm) with the distribution of martensite superimposed in case of a NiTi coil. For the highest stroke, the left and right insets show the distribution of martensitic volume fraction and the equivalent transformation strain computed as $I_2(\bm{H}^{\rm tr})$ from \eqref{eq:convex-set} within an arbitrary cross-section, respectively. The surface point closest to the spring axis is marked with a magenta circle. The distribution of these variables is inhomogeneous, with distinct maxima in the marked point and minima in the central part but below the center of the cross-section. This indicates the predominantly bending character of the loading at this stroke, with a clear asymmetry leading to the shrinking of the region where the material is predominantly compressed. It can also be observed that the phase transformation is not complete in any region of the NiTi coil.

\begin{figure}[h]
\centering
\includegraphics[width=0.95\textwidth]{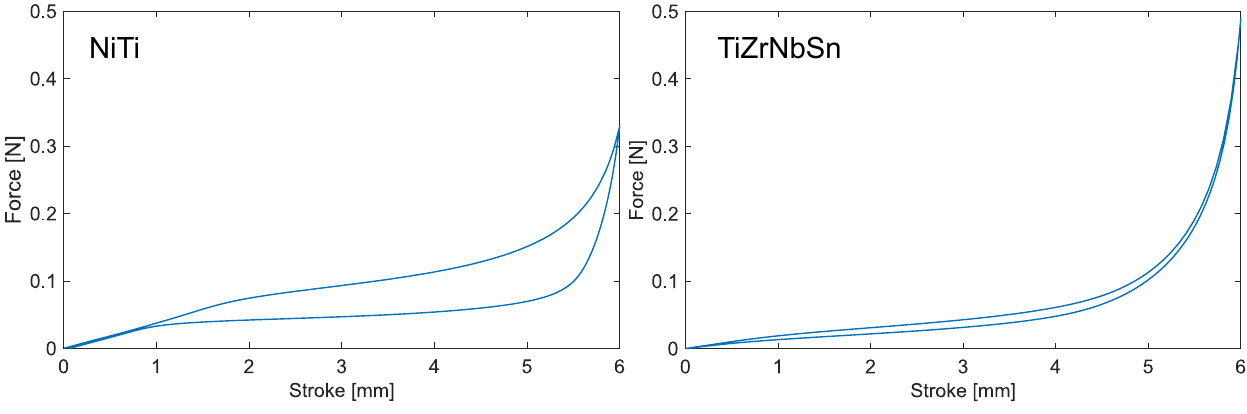}
\caption{The evolution of force with stroke from finite-element simulations of NiTi (left) and Ti-18Zr-11Nb-3Sn (right) coils.}\label{fig:coils_F-u}
\end{figure}

The same computational outputs are provided for a Ti-18Zr-11Nb-3Sn coil in Fig.~\ref{fig:coil_TiNbZrSn_distrib}. In contrast, the phase transformation is completed in a significant volume of the material, and the torsional character of the loading mode is more apparent. This is consistent with much lower attainable transformation strains compared to NiTi and (the assumed) tension-compression symmetry of the response. Finally, Fig.~\ref{fig:coils_F-u} compares the computed macroscopic stroke-force response for both materials in a complete stretching/unstretching cycle. The lower dissipation energy within a superelastic cycle, as exhibited in Fig.~\ref{fig:mater_stress-strain}, results in a narrower hysteresis loop for the Ti-18Zr-11Nb-3Sn alloy. In the parts where the transformation and reorientation are completed, the material deforms only elastically, which leads to a much steeper increase in force than for NiTi.

\section{Conclusions}
\label{sec:conclu}

We have extended a small-strain model based on \cite{SED-FRO-IJOP} to the finite-strain regime using the consistent modular approach of \cite{MIE-LAM-02}. The approach offers an alternative to both the ``classical" multiplicative decomposition-based approach, which necessitates the complete rebuilding of the model structure and comprehensive mathematical manipulations, and to the logarithmic stress rate approach, which builds upon the original model but requires substantial computational effort linked with logarithmic co-rotational integration. 
The logarithmic strain approach employed in this work preserves the original model structure, redirecting the modeling effort to the pre- and post-processing phases, with spectral decomposition of tensors being the primary computational burden. The modular structure of the logarithmic strain approach is also extremely beneficial for future modifications/development, e.g., towards capturing the plastic processes in NiTi martensite, cf. \cite{KWINKS}, or their coupling with phase transformation \cite{HEL-SED-19}. It is worth noting that the relative simplicity of implementation is counterbalanced by limited applicability in particularly severe loading modes; see \cite{FRI-STE-22} for details.

The free energy function has been refined (with respect to the original small-strain model) by incorporating an internal energy term to enhance the flexibility of the transformation kinetics description, as demonstrated in two examples of real material responses. The model has been implemented using the finite element method with the constitutive state update procedure retained from the original implementation, which substantially reduces the development time. Two computational simulations of simple mechanical structures, a cantilever beam and a helical spring, have been conducted for two different SMA: NiTi and Ti-18Zr-11Nb-3Sn. The model plausibly reproduced the shape changes associated with large rotations, which is particularly useful for simulations of slender structures used in microelectromechanical systems or medical devices. The simulations also demonstrated that it captures not only the differences in macroscopic mechanical response but also in the distributions of phase and strain states stemming from the different material responses. Such information is valuable for the development and design of applications relying on SMA alternatives to NiTi. We note, however, that the present results on Ti-18Zr-11Nb-3Sn are merely illustrative, as the experimental input required for model fitting is incomplete, and the model premises beyond the superelastic regime have not yet been experimentally validated. 

\backmatter

\bmhead{Acknowledgements}

This work has been financially supported by the Czech Science Foundation [project No. 24-10366S] and by the Operational Programme Johannes Amos Comenius of the Ministry of Education, Youth and Sport of the Czech Republic, within the frame of project Ferroic Multifunctionalities (FerrMion) [project No. CZ.02.01.01/00/22\_008/0004591], co-funded by the European Union.

\bmhead{Data Availability} Data can be found at: https://doi.org/10.5281/zenodo.15396379.

\bibliography{liter-2025}


\begin{thebibliography}{66}
\ifx \bisbn   \undefined \def \bisbn  #1{ISBN #1}\fi
\ifx \binits  \undefined \def \binits#1{#1}\fi
\ifx \bauthor  \undefined \def \bauthor#1{#1}\fi
\ifx \batitle  \undefined \def \batitle#1{#1}\fi
\ifx \bjtitle  \undefined \def \bjtitle#1{#1}\fi
\ifx \bvolume  \undefined \def \bvolume#1{\textbf{#1}}\fi
\ifx \byear  \undefined \def \byear#1{#1}\fi
\ifx \bissue  \undefined \def \bissue#1{#1}\fi
\ifx \bfpage  \undefined \def \bfpage#1{#1}\fi
\ifx \blpage  \undefined \def \blpage #1{#1}\fi
\ifx \burl  \undefined \def \burl#1{\textsf{#1}}\fi
\ifx \doiurl  \undefined \def \doiurl#1{\url{https://doi.org/#1}}\fi
\ifx \betal  \undefined \def \betal{\textit{et al.}}\fi
\ifx \binstitute  \undefined \def \binstitute#1{#1}\fi
\ifx \binstitutionaled  \undefined \def \binstitutionaled#1{#1}\fi
\ifx \bctitle  \undefined \def \bctitle#1{#1}\fi
\ifx \beditor  \undefined \def \beditor#1{#1}\fi
\ifx \bpublisher  \undefined \def \bpublisher#1{#1}\fi
\ifx \bbtitle  \undefined \def \bbtitle#1{#1}\fi
\ifx \bedition  \undefined \def \bedition#1{#1}\fi
\ifx \bseriesno  \undefined \def \bseriesno#1{#1}\fi
\ifx \blocation  \undefined \def \blocation#1{#1}\fi
\ifx \bsertitle  \undefined \def \bsertitle#1{#1}\fi
\ifx \bsnm \undefined \def \bsnm#1{#1}\fi
\ifx \bsuffix \undefined \def \bsuffix#1{#1}\fi
\ifx \bparticle \undefined \def \bparticle#1{#1}\fi
\ifx \barticle \undefined \def \barticle#1{#1}\fi
\bibcommenthead
\ifx \bconfdate \undefined \def \bconfdate #1{#1}\fi
\ifx \botherref \undefined \def \botherref #1{#1}\fi
\ifx \url \undefined \def \url#1{\textsf{#1}}\fi
\ifx \bchapter \undefined \def \bchapter#1{#1}\fi
\ifx \bbook \undefined \def \bbook#1{#1}\fi
\ifx \bcomment \undefined \def \bcomment#1{#1}\fi
\ifx \oauthor \undefined \def \oauthor#1{#1}\fi
\ifx \citeauthoryear \undefined \def \citeauthoryear#1{#1}\fi
\ifx \endbibitem  \undefined \def \endbibitem {}\fi
\ifx \bconflocation  \undefined \def \bconflocation#1{#1}\fi
\ifx \arxivurl  \undefined \def \arxivurl#1{\textsf{#1}}\fi
\csname PreBibitemsHook\endcsname

\bibitem[\protect\citeauthoryear{Gierig et~al.}{2025}]{Gierig2025}
\begin{barticle}
\bauthor{\bsnm{Gierig}, \binits{M.}},
\bauthor{\bsnm{Liu}, \binits{F.}},
\bauthor{\bsnm{Junker}, \binits{P.}}:
\batitle{Nitinol stent placement in a stenosed artery: A highly nonlinear application scenario for two novel finite-element models}.
\bjtitle{Shape Memory and Superelasticity}
\bvolume{11}(\bissue{1}),
\bfpage{19}--\blpage{33}
(\byear{2025})
\doiurl{10.1007/s40830-025-00525-0}
\end{barticle}
\endbibitem

\bibitem[\protect\citeauthoryear{Mohd~Jani et~al.}{2017}]{JAN-SUB-17}
\begin{barticle}
\bauthor{\bsnm{Mohd~Jani}, \binits{J.}},
\bauthor{\bsnm{Leary}, \binits{M.}},
\bauthor{\bsnm{Subic}, \binits{A.}}:
\batitle{Designing shape memory alloy linear actuators: A review}.
\bjtitle{J. Intel. Mat. Syst. Str.}
\bvolume{28},
\bfpage{1699}--\blpage{1718}
(\byear{2017})
\doiurl{10.1177/1045389X166792}
\end{barticle}
\endbibitem

\bibitem[\protect\citeauthoryear{Naghipour et~al.}{2024}]{Naghipour2024}
\begin{barticle}
\bauthor{\bsnm{Naghipour}, \binits{P.}},
\bauthor{\bsnm{Padula}, \binits{S.}},
\bauthor{\bsnm{Creager}, \binits{C.}},
\bauthor{\bsnm{Oravec}, \binits{H.}}:
\batitle{Large-scale numerical models for shape memory mars spring tires: Development and implementation}.
\bjtitle{Shape Memory and Superelasticity}
\bvolume{10}(\bissue{3}),
\bfpage{341}--\blpage{355}
(\byear{2024})
\doiurl{10.1007/s40830-024-00501-0}
\end{barticle}
\endbibitem

\bibitem[\protect\citeauthoryear{Abavisani et~al.}{2021}]{Abavisani2021}
\begin{barticle}
\bauthor{\bsnm{Abavisani}, \binits{I.}},
\bauthor{\bsnm{Rezaifar}, \binits{O.}},
\bauthor{\bsnm{Kheyroddin}, \binits{A.}}:
\batitle{Multifunctional properties of shape memory materials in civil engineering applications: A state-of-the-art review}.
\bjtitle{Journal of Building Engineering}
\bvolume{44},
\bfpage{102657}
(\byear{2021})
\doiurl{10.1016/j.jobe.2021.102657}
\end{barticle}
\endbibitem

\bibitem[\protect\citeauthoryear{Lee}{1969}]{LEE-69}
\begin{barticle}
\bauthor{\bsnm{Lee}, \binits{E.H.}}:
\batitle{Elastic-plastic deformation at finite strains}.
\bjtitle{ASME J. Appl. Mech.}
\bvolume{36},
\bfpage{1}--\blpage{6}
(\byear{1969})
\doiurl{10.1115/1.3564580}
\end{barticle}
\endbibitem

\bibitem[\protect\citeauthoryear{Helm}{2007}]{HEL-07b}
\begin{barticle}
\bauthor{\bsnm{Helm}, \binits{D.}}:
\batitle{Thermomechanics of martensitic phase transitions in shape memory alloysi. constitutive theories for small and large deformations}.
\bjtitle{J. Mech. Mat. Struct.}
\bvolume{2},
\bfpage{87}--\blpage{112}
(\byear{2007})
\doiurl{10.2140/jomms.2007.2.87}
\end{barticle}
\endbibitem

\bibitem[\protect\citeauthoryear{Reese and Christ}{2008}]{REE-CHR}
\begin{barticle}
\bauthor{\bsnm{Reese}, \binits{S.}},
\bauthor{\bsnm{Christ}, \binits{D.}}:
\batitle{Finite deformation pseudo-elasticity of shape memory alloys - {C}onstitutive modelling and finite element implementation}.
\bjtitle{Int. J. Plasticity}
\bvolume{24},
\bfpage{455}--\blpage{482}
(\byear{2008})
\doiurl{10.1016/j.ijplas.2007.05.005}
\end{barticle}
\endbibitem

\bibitem[\protect\citeauthoryear{Evangelista et~al.}{2009}]{EVA-SAC-09}
\begin{barticle}
\bauthor{\bsnm{Evangelista}, \binits{V.}},
\bauthor{\bsnm{Marfia}, \binits{S.}},
\bauthor{\bsnm{Sacco}, \binits{E.}}:
\batitle{A 3d sma constitutive model in the framework of finite strain}.
\bjtitle{International Journal for Numerical Methods in Engineering}
\bvolume{81}(\bissue{6}),
\bfpage{761}--\blpage{785}
(\byear{2009})
\doiurl{10.1002/nme.2717}
\end{barticle}
\endbibitem

\bibitem[\protect\citeauthoryear{Arghavani et~al.}{2010}]{ARG-AUR-10}
\begin{barticle}
\bauthor{\bsnm{Arghavani}, \binits{J.}},
\bauthor{\bsnm{Auricchio}, \binits{F.}},
\bauthor{\bsnm{Naghdabadi}, \binits{R.}},
\bauthor{\bsnm{Reali}, \binits{A.}},
\bauthor{\bsnm{Sohrabpour}, \binits{S.}}:
\batitle{A {3D} finite strain phenomenological constitutive model for shape memory alloys considering martensite reorientation}.
\bjtitle{Continuum Mechanics and Thermodynamics}
\bvolume{22}(\bissue{5}),
\bfpage{345}--\blpage{362}
(\byear{2010})
\doiurl{10.1007/s00161-010-0155-8}
\end{barticle}
\endbibitem

\bibitem[\protect\citeauthoryear{Arghavani et~al.}{2011}]{ARG-AUR-11b}
\begin{barticle}
\bauthor{\bsnm{Arghavani}, \binits{J.}},
\bauthor{\bsnm{Auricchio}, \binits{F.}},
\bauthor{\bsnm{Naghdabadi}, \binits{R.}}:
\batitle{A finite strain kinematic hardening constitutive model based on hencky strain: General framework, solution algorithm and application to shape memory alloys}.
\bjtitle{Int. J. Plasticity}
\bvolume{27},
\bfpage{940}--\blpage{961}
(\byear{2011})
\doiurl{10.1016/j.ijplas.2010.10.006}
\end{barticle}
\endbibitem

\bibitem[\protect\citeauthoryear{Stupkiewicz and Petryk}{2013}]{STU-PET-12}
\begin{barticle}
\bauthor{\bsnm{Stupkiewicz}, \binits{S.}},
\bauthor{\bsnm{Petryk}, \binits{H.}}:
\batitle{A robust model of pseudoelasticity in shape memory alloys}.
\bjtitle{Int. J. Numer. Methods Eng.}
\bvolume{93},
\bfpage{747}--\blpage{769}
(\byear{2013})
\doiurl{10.1002/nme.4405}
\end{barticle}
\endbibitem

\bibitem[\protect\citeauthoryear{Damanpack et~al.}{2017}]{DAM-LIA-17}
\begin{barticle}
\bauthor{\bsnm{Damanpack}, \binits{A.R.}},
\bauthor{\bsnm{Bodaghi}, \binits{M.}},
\bauthor{\bsnm{Liao}, \binits{W.H.}}:
\batitle{A finite-strain constitutive model for anisotropic shape memory alloys}.
\bjtitle{Mech. Mater.}
\bvolume{112},
\bfpage{129}--\blpage{142}
(\byear{2017})
\doiurl{10.1016/j.mechmat.2017.05.012}
\end{barticle}
\endbibitem

\bibitem[\protect\citeauthoryear{Wang et~al.}{2017}]{WAN-ZAK-17}
\begin{barticle}
\bauthor{\bsnm{Wang}, \binits{J.}},
\bauthor{\bsnm{Moumni}, \binits{Z.}},
\bauthor{\bsnm{Zhang}, \binits{W.}},
\bauthor{\bsnm{Xu}, \binits{Y.}},
\bauthor{\bsnm{Zaki}, \binits{W.}}:
\batitle{A 3{D} finite-strain-based constitutive model for shape memory alloys accounting for thermomechanical coupling and martensite reorientation}.
\bjtitle{Smart Mater. Struct.}
\bvolume{26},
\bfpage{065006}
(\byear{2017})
\doiurl{10.1088/1361-665X/aa6c17}
\end{barticle}
\endbibitem

\bibitem[\protect\citeauthoryear{Christ and Reese}{2009}]{CHR-REE-09}
\begin{barticle}
\bauthor{\bsnm{Christ}, \binits{D.}},
\bauthor{\bsnm{Reese}, \binits{S.}}:
\batitle{A finite element model for shape memory alloys considering thermomechanical couplings at large strains}.
\bjtitle{International Journal of Solids and Structures}
\bvolume{46}(\bissue{20}),
\bfpage{3694}--\blpage{3709}
(\byear{2009})
\doiurl{10.1016/j.ijsolstr.2009.06.017}
\end{barticle}
\endbibitem

\bibitem[\protect\citeauthoryear{Sielenk\"{a}mper and Wulfinghoff}{2022}]{SIE-WUL-22}
\begin{barticle}
\bauthor{\bsnm{Sielenk\"{a}mper}, \binits{M.}},
\bauthor{\bsnm{Wulfinghoff}, \binits{S.}}:
\batitle{A thermomechanical finite strain shape memory alloy model and its application to bistable actuators}.
\bjtitle{Acta Mech.}
\bvolume{3059--3094},
\bfpage{233}
(\byear{2022})
\doiurl{10.1007/s00707-022-03236-0}
\end{barticle}
\endbibitem

\bibitem[\protect\citeauthoryear{M\"{u}ller and Bruhns}{2006}]{MUL-BRU-06}
\begin{barticle}
\bauthor{\bsnm{M\"{u}ller}, \binits{C.}},
\bauthor{\bsnm{Bruhns}, \binits{O.T.}}:
\batitle{A thermodynamic finite-strain model for pseudoelastic shape memory alloys}.
\bjtitle{Int. J. Plasticity}
\bvolume{22},
\bfpage{1658}--\blpage{1682}
(\byear{2006})
\doiurl{10.1016/j.ijplas.2006.02.010}
\end{barticle}
\endbibitem

\bibitem[\protect\citeauthoryear{Xu et~al.}{2019}]{XU-LAG-19}
\begin{barticle}
\bauthor{\bsnm{Xu}, \binits{L.}},
\bauthor{\bsnm{Baxevanis}, \binits{T.}},
\bauthor{\bsnm{Lagoudas}, \binits{D.}}:
\batitle{A three-dimensional constitutive model for the martensitic transformation in polycrystalline shape memory alloys under large deformation}.
\bjtitle{Smart Mater. Struct.}
\bvolume{28},
\bfpage{074004}
(\byear{2019})
\doiurl{10.1088/1361-665X/ab1acb}
\end{barticle}
\endbibitem

\bibitem[\protect\citeauthoryear{Xu et~al.}{2021}]{XU-LAG-21}
\begin{barticle}
\bauthor{\bsnm{Xu}, \binits{L.}},
\bauthor{\bsnm{Solomou}, \binits{A.}},
\bauthor{\bsnm{Baxevanis}, \binits{T.}},
\bauthor{\bsnm{Lagoudas}, \binits{D.}}:
\batitle{Finite strain constitutive modeling for shape memory alloys considering transformation-induced plasticity and two-way shape memory effect}.
\bjtitle{International Journal of Solids and Structures}
\bvolume{221},
\bfpage{42}--\blpage{59}
(\byear{2021})
\doiurl{10.1016/j.ijsolstr.2020.03.009}
\end{barticle}
\endbibitem

\bibitem[\protect\citeauthoryear{Zhang and Baxevanis}{2021}]{ZHA-BAX-21}
\begin{botherref}
\oauthor{\bsnm{Zhang}, \binits{M.}},
\oauthor{\bsnm{Baxevanis}, \binits{T.}}:
An extended three-dimensional finite strain constitutive model for shape memory alloys.
Journal of Applied Mechanics
\textbf{88}(11)
(2021)
\doiurl{10.1115/1.4051833}
\end{botherref}
\endbibitem

\bibitem[\protect\citeauthoryear{Xiao et~al.}{1997}]{XIA-BRU-97}
\begin{barticle}
\bauthor{\bsnm{Xiao}, \binits{H.}},
\bauthor{\bsnm{Bruhns}, \binits{O.T.}},
\bauthor{\bsnm{Meyers}, \binits{A.}}:
\batitle{Logarithmic strain, logarithmic spin and logarithmic rate}.
\bjtitle{Acta Mechanica}
\bvolume{124}(\bissue{1-4}),
\bfpage{89}--\blpage{105}
(\byear{1997})
\doiurl{10.1007/bf01213020}
\end{barticle}
\endbibitem

\bibitem[\protect\citeauthoryear{Miehe et~al.}{2002}]{MIE-LAM-02}
\begin{barticle}
\bauthor{\bsnm{Miehe}, \binits{C.}},
\bauthor{\bsnm{Apel}, \binits{N.}},
\bauthor{\bsnm{Lambrecht}, \binits{M.}}:
\batitle{Anisotropic additive plasticity in the logarithmic strain space: modular kinematic formulation and implementation based on incremental minimization principles for standard materials}.
\bjtitle{Computer Methods in Applied Mechanics and Engineering}
\bvolume{191}(\bissue{47--48}),
\bfpage{5383}--\blpage{5425}
(\byear{2002})
\doiurl{10.1016/s0045-7825(02)00438-3}
\end{barticle}
\endbibitem

\bibitem[\protect\citeauthoryear{Aldakheel}{2017}]{Aldakheel2017}
\begin{barticle}
\bauthor{\bsnm{Aldakheel}, \binits{F.}}:
\batitle{Micromorphic approach for gradient-extended thermo-elastic-plastic solids in the logarithmic strain space}.
\bjtitle{Continuum Mechanics and Thermodynamics}
\bvolume{29}(\bissue{6}),
\bfpage{1207}--\blpage{1217}
(\byear{2017})
\doiurl{10.1007/s00161-017-0571-0}
\end{barticle}
\endbibitem

\bibitem[\protect\citeauthoryear{Hor\'{a}k and Jir\'{a}sek}{2013}]{Horak2013}
\begin{bchapter}
\bauthor{\bsnm{Hor\'{a}k}, \binits{M.}},
\bauthor{\bsnm{Jir\'{a}sek}, \binits{M.}}:
\bctitle{An extension of small-strain models to the large-strain range based on an additive decomposition of a logarithmic strain}.
In: \beditor{\bsnm{Chleboun}, \binits{J.}},
\beditor{\bsnm{Segeth}, \binits{K.}},
\beditor{\bsnm{\v{S}\'{i}stek}, \binits{J.}},
\beditor{\bsnm{Vejchodsk\'{y}}, \binits{T.}} (eds.)
\bbtitle{Programs and Algorithms of Numerical Mathematics. Proceedings of Seminar.},
pp. \bfpage{88}--\blpage{93}
(\byear{2013}).
\bcomment{Institute of Mathematics AS CR, Prague}
\end{bchapter}
\endbibitem

\bibitem[\protect\citeauthoryear{Holthusen et~al.}{2022}]{Holthusen2022}
\begin{barticle}
\bauthor{\bsnm{Holthusen}, \binits{H.}},
\bauthor{\bsnm{Brepols}, \binits{T.}},
\bauthor{\bsnm{Reese}, \binits{S.}},
\bauthor{\bsnm{Simon}, \binits{J.-W.}}:
\batitle{A two-surface gradient-extended anisotropic damage model using a second order damage tensor coupled to additive plasticity in the logarithmic strain space}.
\bjtitle{Journal of the Mechanics and Physics of Solids}
\bvolume{163},
\bfpage{104833}
(\byear{2022})
\doiurl{10.1016/j.jmps.2022.104833}
\end{barticle}
\endbibitem

\bibitem[\protect\citeauthoryear{Miehe et~al.}{2011}]{Miehe2011}
\begin{barticle}
\bauthor{\bsnm{Miehe}, \binits{C.}},
\bauthor{\bsnm{Méndez~Diez}, \binits{J.}},
\bauthor{\bsnm{Göktepe}, \binits{S.}},
\bauthor{\bsnm{Schänzel}, \binits{L.-M.}}:
\batitle{Coupled thermoviscoplasticity of glassy polymers in the logarithmic strain space based on the free volume theory}.
\bjtitle{International Journal of Solids and Structures}
\bvolume{48}(\bissue{13}),
\bfpage{1799}--\blpage{1817}
(\byear{2011})
\doiurl{10.1016/j.ijsolstr.2011.01.030}
\end{barticle}
\endbibitem

\bibitem[\protect\citeauthoryear{Friedlein et~al.}{2022}]{FRI-STE-22}
\begin{barticle}
\bauthor{\bsnm{Friedlein}, \binits{J.}},
\bauthor{\bsnm{Mergheim}, \binits{J.}},
\bauthor{\bsnm{Steinmann}, \binits{P.}}:
\batitle{Observations on additive plasticity in the logarithmic strain space at excessive strains}.
\bjtitle{International Journal of Solids and Structures}
\bvolume{239--240},
\bfpage{111416}
(\byear{2022})
\doiurl{10.1016/j.ijsolstr.2021.111416}
\end{barticle}
\endbibitem

\bibitem[\protect\citeauthoryear{Woodworth and Kaliske}{2024}]{WOO-KAL-24}
\begin{barticle}
\bauthor{\bsnm{Woodworth}, \binits{L.A.}},
\bauthor{\bsnm{Kaliske}, \binits{M.}}:
\batitle{Damage in a comprehensive model for shape memory alloys in logarithmic strain space}.
\bjtitle{Computer Methods in Applied Mechanics and Engineering}
\bvolume{421},
\bfpage{116769}
(\byear{2024})
\doiurl{10.1016/j.cma.2024.116769}
\end{barticle}
\endbibitem

\bibitem[\protect\citeauthoryear{Chattopadhyay et~al.}{2024}]{CHA-KAL-24}
\begin{botherref}
\oauthor{\bsnm{Chattopadhyay}, \binits{S.}},
\oauthor{\bsnm{Woodworth}, \binits{L.A.}},
\oauthor{\bsnm{Kaliske}, \binits{M.}}:
Finite strain modelling of shape memory alloys in the logarithmic strain space: A comparative study with other finite strain approaches.
Int J. Solids. Struct.
\textbf{299}(112892)
(2024)
\doiurl{10.2139/ssrn.4766223}
\end{botherref}
\endbibitem

\bibitem[\protect\citeauthoryear{Souza et~al.}{1998}]{SOU-98}
\begin{barticle}
\bauthor{\bsnm{Souza}, \binits{A.C.}},
\bauthor{\bsnm{Mamiya}, \binits{E.N.}},
\bauthor{\bsnm{Zouain}, \binits{N.}}:
\batitle{Three-dimensional model for solids undergoing stress-induced phase transformations}.
\bjtitle{Eur. J. Mech. A}
\bvolume{17},
\bfpage{789}--\blpage{806}
(\byear{1998})
\doiurl{10.1016/S0997-7538(98)80005-3}
\end{barticle}
\endbibitem

\bibitem[\protect\citeauthoryear{Woodworth et~al.}{2022}]{WOO-KAL-22}
\begin{barticle}
\bauthor{\bsnm{Woodworth}, \binits{L.A.}},
\bauthor{\bsnm{Wang}, \binits{X.}},
\bauthor{\bsnm{Lin}, \binits{G.}},
\bauthor{\bsnm{Kaliske}, \binits{M.}}:
\batitle{A multi-featured shape memory alloy constitutive model incorporating tension-compression asymmetric interpolation}.
\bjtitle{Mechanics of Materials}
\bvolume{172},
\bfpage{104392}
(\byear{2022})
\doiurl{10.1016/j.mechmat.2022.104392}
\end{barticle}
\endbibitem

\bibitem[\protect\citeauthoryear{Sedl\'{a}k et~al.}{2012}]{SED-FRO-IJOP}
\begin{barticle}
\bauthor{\bsnm{Sedl\'{a}k}, \binits{P.}},
\bauthor{\bsnm{Frost}, \binits{M.}},
\bauthor{\bsnm{Bene{\v s}ov{\' a}}, \binits{B.}},
\bauthor{\bsnm{{\v S}ittner}, \binits{P.}},
\bauthor{\bsnm{Ben~Zineb}, \binits{T.}}:
\batitle{Thermomechanical model for {N}i{T}i-based shape memory alloys including {R}-phase and material anisotropy under multi-axial loadings}.
\bjtitle{Int. J. Plasticity}
\bvolume{39},
\bfpage{132}--\blpage{151}
(\byear{2012})
\doiurl{10.1016/j.ijplas.2012.06.008}
\end{barticle}
\endbibitem

\bibitem[\protect\citeauthoryear{Frost et~al.}{2016}]{FRO-BEN-MMS}
\begin{barticle}
\bauthor{\bsnm{Frost}, \binits{M.}},
\bauthor{\bsnm{Bene{\v{s}}ov{\'a}}, \binits{B.}},
\bauthor{\bsnm{Sedl{\'a}k}, \binits{P.}}:
\batitle{A microscopically motivated constitutive model for shape memory alloys: formulation, analysis and computations}.
\bjtitle{Math. Mech. Solids}
\bvolume{21}(\bissue{3}),
\bfpage{358}--\blpage{382}
(\byear{2016})
\doiurl{10.1177/1081286514522474}
\end{barticle}
\endbibitem

\bibitem[\protect\citeauthoryear{Halphen and Nguyen}{1975}]{HAL-NGU}
\begin{barticle}
\bauthor{\bsnm{Halphen}, \binits{B.}},
\bauthor{\bsnm{Nguyen}, \binits{Q.S.}}:
\batitle{Sur les mat\'{e}riaux standard g\'{e}n\'{e}ralis\'{e}s}.
\bjtitle{J. Mecanique}
\bvolume{14},
\bfpage{39}--\blpage{63}
(\byear{1975})
\end{barticle}
\endbibitem

\bibitem[\protect\citeauthoryear{Liu and Favier}{2000}]{LIU-FAV}
\begin{barticle}
\bauthor{\bsnm{Liu}, \binits{Y.}},
\bauthor{\bsnm{Favier}, \binits{D.}}:
\batitle{Stabilisation of martensite due to shear deformation via variant reorientation in polycrystalline {NiTi}}.
\bjtitle{Acta Mater.}
\bvolume{48},
\bfpage{3489}--\blpage{3499}
(\byear{2000})
\doiurl{10.1016/S1359-6454(00)00129-4}
\end{barticle}
\endbibitem

\bibitem[\protect\citeauthoryear{Belyaev et~al.}{2020}]{BEL-RES-20}
\begin{barticle}
\bauthor{\bsnm{Belyaev}, \binits{S.}},
\bauthor{\bsnm{Resnina}, \binits{N.}},
\bauthor{\bsnm{Rakhimov}, \binits{T.}},
\bauthor{\bsnm{Andreev}, \binits{V.}}:
\batitle{Martensite stabilisation effect in {Ni}-rich {NiTi} shape memory alloy with different structure and martensitic transformations}.
\bjtitle{Sensors and Actuators A: Physical}
\bvolume{305},
\bfpage{111911}
(\byear{2020})
\doiurl{10.1016/j.sna.2020.111911}
\end{barticle}
\endbibitem

\bibitem[\protect\citeauthoryear{Liu et~al.}{1998}]{LIU-XIE}
\begin{barticle}
\bauthor{\bsnm{Liu}, \binits{Y.}},
\bauthor{\bsnm{Xie}, \binits{Z.}},
\bauthor{\bsnm{Van~Humbeeck}, \binits{J.}}:
\batitle{Asymmetry of stress-strain curves under tension and compression for {NiTi} shape memory alloys}.
\bjtitle{Acta Mater.}
\bvolume{46},
\bfpage{4325}--\blpage{4338}
(\byear{1998})
\doiurl{10.1016/S1359-6454(98)00112-8}
\end{barticle}
\endbibitem

\bibitem[\protect\citeauthoryear{Frost et~al.}{2021a}]{MD-NITIFE}
\begin{barticle}
\bauthor{\bsnm{Frost}, \binits{M.}},
\bauthor{\bsnm{Jury}, \binits{A.}},
\bauthor{\bsnm{Heller}, \binits{L.}},
\bauthor{\bsnm{Sedl\'{a}k}, \binits{P.}}:
\batitle{Experimentally validated constitutive model for {NiTi-based} shape memory alloys featuring intermediate {R-phase} transformation: {A} case study of {Ni48Ti49Fe3}}.
\bjtitle{Mater. Design}
\bvolume{203},
\bfpage{109593}
(\byear{2021})
\doiurl{10.1016/j.matdes.2021.109593}
\end{barticle}
\endbibitem

\bibitem[\protect\citeauthoryear{Frost et~al.}{2021b}]{FRO-BEN-IJSS}
\begin{barticle}
\bauthor{\bsnm{Frost}, \binits{M.}},
\bauthor{\bsnm{Bene\v{s}ov\'a}, \binits{B.}},
\bauthor{\bsnm{Seiner}, \binits{H.}},
\bauthor{\bsnm{Kru\v{z}\'{i}k}, \binits{M.}},
\bauthor{\bsnm{\v{S}ittner}, \binits{P.}},
\bauthor{\bsnm{Sedl\'{a}k}, \binits{P.}}:
\batitle{Thermomechanical model for {NiTi-based} shape memory alloys covering macroscopic localization of martensitic transformation}.
\bjtitle{Int. J. Solids Struct.}
\bvolume{221},
\bfpage{117}--\blpage{129}
(\byear{2021})
\doiurl{10.1016/j.ijsolstr.2020.08.012}
\end{barticle}
\endbibitem

\bibitem[\protect\citeauthoryear{Frost et~al.}{2016}]{JIMSS-SPRING}
\begin{barticle}
\bauthor{\bsnm{Frost}, \binits{M.}},
\bauthor{\bsnm{Sedl{\'a}k}, \binits{P.}},
\bauthor{\bsnm{Kade{\v{r}}{\'a}vek}, \binits{L.}},
\bauthor{\bsnm{Heller}, \binits{L.}},
\bauthor{\bsnm{{\v{S}}ittner}, \binits{P.}}:
\batitle{Modeling of mechanical response of {N}i{T}i shape memory alloy subjected to combined thermal and non-proportional mechanical loading: a case study on helical spring actuator}.
\bjtitle{J. Intel. Mat. Syst. Str.}
\bvolume{27},
\bfpage{1927}--\blpage{1938}
(\byear{2016})
\doiurl{10.1177/1045389X15610908}
\end{barticle}
\endbibitem

\bibitem[\protect\citeauthoryear{Frost et~al.}{2018}]{SMS-SNAKE}
\begin{barticle}
\bauthor{\bsnm{Frost}, \binits{M.}},
\bauthor{\bsnm{Sedl\'{a}k}, \binits{P.}},
\bauthor{\bsnm{Heller}, \binits{L.}},
\bauthor{\bsnm{Kade{\v{r}}{\'a}vek}, \binits{L.}},
\bauthor{\bsnm{\v{S}ittner}, \binits{P.}}:
\batitle{Experimental and computational study on phase transformations in superelastic {N}i{T}i snake-like spring}.
\bjtitle{Smart Mater. Struct.}
\bvolume{27},
\bfpage{095005}
(\byear{2018})
\doiurl{10.1088/1361-665X/aacca4}
\end{barticle}
\endbibitem

\bibitem[\protect\citeauthoryear{Frost et~al.}{2020}]{SMS-DSCT}
\begin{barticle}
\bauthor{\bsnm{Frost}, \binits{M.}},
\bauthor{\bsnm{\v{S}ev\v{c}\'{i}k}, \binits{M.}},
\bauthor{\bsnm{Kade\v{r}\'{a}vek}, \binits{L.}},
\bauthor{\bsnm{\v{S}ittner}, \binits{P.}},
\bauthor{\bsnm{Sedl\'{a}k}, \binits{P.}}:
\batitle{Reconstruction of phase distributions in {NiTi} helical spring: comparison of diffraction/scattering computed tomography and computational modeling}.
\bjtitle{Smart Mater. Struct.}
\bvolume{29},
\bfpage{075036}
(\byear{2020})
\doiurl{10.1088/1361-665X/ab8c26}
\end{barticle}
\endbibitem

\bibitem[\protect\citeauthoryear{Sedlák et~al.}{2021}]{IJSS-DSCT}
\begin{barticle}
\bauthor{\bsnm{Sedlák}, \binits{P.}},
\bauthor{\bsnm{Frost}, \binits{M.}},
\bauthor{\bsnm{\v{S}ev\v{c}\'{i}k}, \binits{M.}},
\bauthor{\bsnm{Seiner}, \binits{H.}}:
\batitle{{3D} spatial reconstruction of macroscopic austenite-martensite transition zones in {NiTi} wires induced by tension and twisting using diffraction/scattering computed tomography}.
\bjtitle{International Journal of Solids and Structures}
\bvolume{228},
\bfpage{111122}
(\byear{2021})
\doiurl{10.1016/j.ijsolstr.2021.111122}
\end{barticle}
\endbibitem

\bibitem[\protect\citeauthoryear{Heller et~al.}{2019}]{HEL-SED-19}
\begin{barticle}
\bauthor{\bsnm{Heller}, \binits{L.}},
\bauthor{\bsnm{\v{S}ittner}, \binits{P.}},
\bauthor{\bsnm{Sedl\'{a}k}, \binits{P.}},
\bauthor{\bsnm{Seiner}, \binits{H.}},
\bauthor{\bsnm{Tyc}, \binits{O.}},
\bauthor{\bsnm{Kade{\v{r}}{\'a}vek}, \binits{L.}},
\bauthor{\bsnm{Sedm{\'a}k}, \binits{P.}},
\bauthor{\bsnm{Vronka}, \binits{M.}}:
\batitle{Beyond the strain recoverability of martensitic transformation in {NiTi}}.
\bjtitle{Int. J. Plasticity}
\bvolume{116},
\bfpage{232}--\blpage{264}
(\byear{2019})
\doiurl{10.1016/j.ijplas.2019.01.007}
\end{barticle}
\endbibitem

\bibitem[\protect\citeauthoryear{Fu et~al.}{2015}]{HOS-MIY-15}
\begin{barticle}
\bauthor{\bsnm{Fu}, \binits{J.}},
\bauthor{\bsnm{Yamamoto}, \binits{A.}},
\bauthor{\bsnm{Kim}, \binits{H.Y.}},
\bauthor{\bsnm{Hosoda}, \binits{H.}},
\bauthor{\bsnm{Miyazaki}, \binits{S.}}:
\batitle{Novel {T}i-base superelastic alloys with large recovery strain and excellent biocompatibility}.
\bjtitle{Acta Biomaterialia}
\bvolume{17},
\bfpage{56}--\blpage{67}
(\byear{2015})
\doiurl{10.1016/j.actbio.2015.02.001}
\end{barticle}
\endbibitem

\bibitem[\protect\citeauthoryear{Köster et~al.}{2000}]{Koester2000}
\begin{barticle}
\bauthor{\bsnm{Köster}, \binits{R.}},
\bauthor{\bsnm{Vieluf}, \binits{D.}},
\bauthor{\bsnm{Kiehn}, \binits{M.}},
\bauthor{\bsnm{Sommerauer}, \binits{M.}},
\bauthor{\bsnm{Kähler}, \binits{J.}},
\bauthor{\bsnm{Baldus}, \binits{S.}},
\bauthor{\bsnm{Meinertz}, \binits{T.}},
\bauthor{\bsnm{Hamm}, \binits{C.W.}}:
\batitle{Nickel and molybdenum contact allergies in patients with coronary in-stent restenosis}.
\bjtitle{The Lancet}
\bvolume{356}(\bissue{9245}),
\bfpage{1895}--\blpage{1897}
(\byear{2000})
\doiurl{10.1016/s0140-6736(00)03262-1}
\end{barticle}
\endbibitem

\bibitem[\protect\citeauthoryear{Miehe and Lambrecht}{2001}]{MIE-LAM-01}
\begin{barticle}
\bauthor{\bsnm{Miehe}, \binits{C.}},
\bauthor{\bsnm{Lambrecht}, \binits{M.}}:
\batitle{Algorithms for computation of stresses and elasticity moduli in terms of {Seth-Hill's} family of generalized strain tensors}.
\bjtitle{Communications in Numerical Methods in Engineering}
\bvolume{17}(\bissue{5}),
\bfpage{337}--\blpage{353}
(\byear{2001})
\doiurl{10.1002/cnm.404}
\end{barticle}
\endbibitem

\bibitem[\protect\citeauthoryear{Bene\v{s}ov\'a et~al.}{2021}]{BEN-FRO-DCDSS}
\begin{barticle}
\bauthor{\bsnm{Bene\v{s}ov\'a}, \binits{B.}},
\bauthor{\bsnm{Frost}, \binits{M.}},
\bauthor{\bsnm{Kade{\v{r}}{\'a}vek}, \binits{L.}},
\bauthor{\bsnm{Roub\'{i}\v{c}ek}, \binits{T.}},
\bauthor{\bsnm{Sedl\'{a}k}, \binits{P.}}:
\batitle{An experimentallly-fitted thermodynamical constitutive model for polycrystalline shape memory alloys}.
\bjtitle{Disc. Cont. Dynam. Syst. S}
\bvolume{14}(\bissue{11}),
\bfpage{3925}--\blpage{3952}
(\byear{2021})
\doiurl{3925-3952}
\end{barticle}
\endbibitem

\bibitem[\protect\citeauthoryear{Sedl\'{a}k et~al.}{2014}]{JMEP-SPRING}
\begin{barticle}
\bauthor{\bsnm{Sedl\'{a}k}, \binits{P.}},
\bauthor{\bsnm{Frost}, \binits{M.}},
\bauthor{\bsnm{Kruisov\'{a}}, \binits{A.}},
\bauthor{\bsnm{Hi\v{r}manov\'{a}}, \binits{K.}},
\bauthor{\bsnm{Heller}, \binits{L.}},
\bauthor{\bsnm{\v{S}ittner}, \binits{P.}}:
\batitle{Simulations of mechanical response of superelastic {N}i{T}i helical spring and its relation to fatigue resistance}.
\bjtitle{J. Mater. Eng. Perform.}
\bvolume{23},
\bfpage{2591}--\blpage{2598}
(\byear{2014})
\doiurl{10.1007/s11665-014-0906-y}
\end{barticle}
\endbibitem

\bibitem[\protect\citeauthoryear{Frost et~al.}{2014}]{JMEP-STENT}
\begin{barticle}
\bauthor{\bsnm{Frost}, \binits{M.}},
\bauthor{\bsnm{Sedl\'{a}k}, \binits{P.}},
\bauthor{\bsnm{Kruisov\'{a}}, \binits{A.}},
\bauthor{\bsnm{Landa}, \binits{M.}}:
\batitle{Simulations of self-expanding braided stent using macroscopic model of {N}i{T}i shape memory alloys covering {R}-phase}.
\bjtitle{J. Mater. Eng. Perform.}
\bvolume{23},
\bfpage{2584}--\blpage{2590}
(\byear{2014})
\doiurl{10.1007/s11665-014-0966-z}
\end{barticle}
\endbibitem

\bibitem[\protect\citeauthoryear{Chemisky et~al.}{2011}]{CHE-DUV}
\begin{barticle}
\bauthor{\bsnm{Chemisky}, \binits{Y.}},
\bauthor{\bsnm{Duval}, \binits{A.}},
\bauthor{\bsnm{Patoor}, \binits{E.}},
\bauthor{\bsnm{Ben~Zineb}, \binits{T.}}:
\batitle{Constitutive model for shape memory alloys including phase transformation, martensitic reorientation and twins accommodation}.
\bjtitle{Mech. Mater.}
\bvolume{43},
\bfpage{361}--\blpage{376}
(\byear{2011})
\doiurl{10.1016/j.mechmat.2011.04.003}
\end{barticle}
\endbibitem

\bibitem[\protect\citeauthoryear{Lagoudas et~al.}{2012}]{LAG-CHE}
\begin{barticle}
\bauthor{\bsnm{Lagoudas}, \binits{D.C.}},
\bauthor{\bsnm{Hartl}, \binits{D.J.}},
\bauthor{\bsnm{Chemisky}, \binits{Y.}},
\bauthor{\bsnm{Machado}, \binits{L.G.}},
\bauthor{\bsnm{Popov}, \binits{P.}}:
\batitle{Constitutive model for the numerical analysis of phase transformation in polycrystalline shape memory alloys}.
\bjtitle{Int. J. Plasticity}
\bvolume{32--33},
\bfpage{155}--\blpage{183}
(\byear{2012})
\doiurl{10.1016/j.ijplas.2011.10.009}
\end{barticle}
\endbibitem

\bibitem[\protect\citeauthoryear{Lagoudas et~al.}{2006}]{PAT-LAG-II}
\begin{barticle}
\bauthor{\bsnm{Lagoudas}, \binits{D.C.}},
\bauthor{\bsnm{Entchev}, \binits{P.B.}},
\bauthor{\bsnm{Popov}, \binits{P.}},
\bauthor{\bsnm{Patoor}, \binits{E.}},
\bauthor{\bsnm{Brinson}, \binits{L.C.}},
\bauthor{\bsnm{Gao}, \binits{X.}}:
\batitle{Shape memory alloys, {P}art {II}: {M}odeling of polycrystals}.
\bjtitle{Mech. Mater.}
\bvolume{38},
\bfpage{430}--\blpage{62}
(\byear{2006})
\doiurl{10.1016/j.mechmat.2005.08.003}
\end{barticle}
\endbibitem

\bibitem[\protect\citeauthoryear{Frost and Valdman}{2022}]{FRO-VAL-22}
\begin{barticle}
\bauthor{\bsnm{Frost}, \binits{M.}},
\bauthor{\bsnm{Valdman}, \binits{J.}}:
\batitle{Vectorized {MATLAB} implementation of the incremental minimization principle for rate-independent dissipative solids using {FEM}: {A} constitutive model of shape memory alloys}.
\bjtitle{Mathematics}
\bvolume{10}(\bissue{23}),
\bfpage{4412}
(\byear{2022})
\doiurl{10.3390/math10234412}
\end{barticle}
\endbibitem

\bibitem[\protect\citeauthoryear{Hackl and Fischer}{2008}]{HAC-FIS}
\begin{barticle}
\bauthor{\bsnm{Hackl}, \binits{K.}},
\bauthor{\bsnm{Fischer}, \binits{F.D.}}:
\batitle{On the relation between the principle of maximum dissipation and inelastic evolution given by dissipation potentials}.
\bjtitle{Proc. R. Soc. London, Ser. A}
\bvolume{464},
\bfpage{117}--\blpage{132}
(\byear{2008})
\doiurl{10.1098/rspa.2007.0086}
\end{barticle}
\endbibitem

\bibitem[\protect\citeauthoryear{{\v C}erm\'{a}k et~al.}{2019}]{CSV-19}
\begin{barticle}
\bauthor{\bsnm{{\v C}erm\'{a}k}, \binits{M.}},
\bauthor{\bsnm{Sysala}, \binits{S.}},
\bauthor{\bsnm{Valdman}, \binits{J.}}:
\batitle{Efficient and flexible {MATLAB} implementation of {2D} and {3D} elastoplastic problems}.
\bjtitle{Appl. Math. Comput.}
\bvolume{355},
\bfpage{595}--\blpage{614}
(\byear{2019})
\doiurl{10.1016/j.amc.2019.02.054}
\end{barticle}
\endbibitem

\bibitem[\protect\citeauthoryear{Posp\'{i}\v{s}il et~al.}{2023}]{Pospisil2023}
\begin{bchapter}
\bauthor{\bsnm{Posp\'{i}\v{s}il}, \binits{L.}},
\bauthor{\bsnm{Sysala}, \binits{S.}},
\bauthor{\bsnm{\v{C}erm\'{a}k}, \binits{M.}}:
\bctitle{Spectral projected gradient method for conic optimization in kinematic limit analysis}.
In: \bbtitle{International Conference of Numerical Analysis and Applied Mathematics: ICNAAM 2021},
vol. \bseriesno{2849},
p. \bfpage{310007}.
\bpublisher{AIP Publishing},
\blocation{Melville: USA}
(\byear{2023}).
\doiurl{10.1063/5.0162195}
\end{bchapter}
\endbibitem

\bibitem[\protect\citeauthoryear{Sv\v{e}tl\'{i}k et~al.}{2024}]{Svetlik2024}
\begin{bchapter}
\bauthor{\bsnm{Sv\v{e}tl\'{i}k}, \binits{T.}},
\bauthor{\bsnm{Varga}, \binits{R.}},
\bauthor{\bsnm{Posp\'{i}\v{s}il}, \binits{L.}},
\bauthor{\bsnm{\v{C}erm\'{a}k}, \binits{M.}}:
\bctitle{Interface between {A}nsys and {M}atlab for solving elastic problems with non-conforming meshes}.
In: \bbtitle{International Conference of Numerical Analysis and Applied Mathematics: ICNAAM 2022},
vol. \bseriesno{3094},
p. \bfpage{300005}.
\bpublisher{AIP Publishing},
\blocation{Melville: USA}
(\byear{2024}).
\doiurl{10.1063/5.0210891}
\end{bchapter}
\endbibitem

\bibitem[\protect\citeauthoryear{Kar\'{a}tson et~al.}{2025}]{Karatson2025}
\begin{barticle}
\bauthor{\bsnm{Kar\'{a}tson}, \binits{J.}},
\bauthor{\bsnm{Sysala}, \binits{S.}},
\bauthor{\bsnm{B\'{e}re\v{s}}, \binits{M.}}:
\batitle{Quasi-{N}ewton iterative solution approaches for nonsmooth elliptic operators with applications to elasto-plasticity}.
\bjtitle{Computers and Mathematics with Applications}
\bvolume{178},
\bfpage{61}--\blpage{80}
(\byear{2025})
\doiurl{10.1016/j.camwa.2024.11.022}
\end{barticle}
\endbibitem

\bibitem[\protect\citeauthoryear{Weinstein and Rao}{2017}]{ADiGator}
\begin{barticle}
\bauthor{\bsnm{Weinstein}, \binits{M.J.}},
\bauthor{\bsnm{Rao}, \binits{A.V.}}:
\batitle{Algorithm 984: Adigator, a toolbox for the algorithmic differentiation of mathematical functions in matlab using source transformation via operator overloading}.
\bjtitle{ACM Transactions on Mathematical Software}
\bvolume{44}(\bissue{2}),
\bfpage{1}--\blpage{25}
(\byear{2017})
\doiurl{10.1145/3104990}
\end{barticle}
\endbibitem

\bibitem[\protect\citeauthoryear{Moskovka et~al.}{in preparation}]{Moskovka2025}
\begin{botherref}
\oauthor{\bsnm{Moskovka}, \binits{A.}},
\oauthor{\bsnm{Frost}, \binits{M.}},
\oauthor{\bsnm{Valdman}, \binits{J.}}:
Efficient matlab finite element implementation of rate-independent material models in finite strain
(in preparation)
\end{botherref}
\endbibitem

\bibitem[\protect\citeauthoryear{{Fort Wayne Metals Corp.}}{}]{FWM}
\begin{botherref}
\oauthor{\bsnm{{Fort Wayne Metals Corp.}}}:
https://www.fwmetals.com/what-we-do/materials/nitinol/superelastic-nitinol.
Last accessed: 31/3/2025
\end{botherref}
\endbibitem

\bibitem[\protect\citeauthoryear{Jia et~al.}{2024}]{JIA-GLO-24}
\begin{barticle}
\bauthor{\bsnm{Jia}, \binits{T.}},
\bauthor{\bsnm{Guines}, \binits{D.}},
\bauthor{\bsnm{Laillé}, \binits{D.}},
\bauthor{\bsnm{Leotoing}, \binits{L.}},
\bauthor{\bsnm{Gloriant}, \binits{T.}}:
\batitle{Finite element analysis of the mechanical performance of self-expanding endovascular stents made with new nickel-free superelastic {$\beta$}-titanium alloys}.
\bjtitle{Journal of the Mechanical Behavior of Biomedical Materials}
\bvolume{151},
\bfpage{106345}
(\byear{2024})
\doiurl{10.1016/j.jmbbm.2023.106345}
\end{barticle}
\endbibitem

\bibitem[\protect\citeauthoryear{Sedm{\'a}k et~al.}{2016}]{SCIENCE}
\begin{barticle}
\bauthor{\bsnm{Sedm{\'a}k}, \binits{P.}},
\bauthor{\bsnm{Pilch}, \binits{J.}},
\bauthor{\bsnm{Heller}, \binits{L.}},
\bauthor{\bsnm{Kope{\v{c}}ek}, \binits{J.}},
\bauthor{\bsnm{Wright}, \binits{J.}},
\bauthor{\bsnm{Sedl{\'a}k}, \binits{P.}},
\bauthor{\bsnm{Frost}, \binits{M.}},
\bauthor{\bsnm{{\v{S}}ittner}, \binits{P.}}:
\batitle{Grain-resolved analysis of localized deformation in nickel-titanium wire under tensile load}.
\bjtitle{Science}
\bvolume{353}(\bissue{6299}),
\bfpage{559}--\blpage{562}
(\byear{2016})
\doiurl{10.1126/science.aad6700}
\end{barticle}
\endbibitem

\bibitem[\protect\citeauthoryear{Kim and Miyazaki}{2018}]{KIM-MIY-18}
\begin{bbook}
\bauthor{\bsnm{Kim}, \binits{H.Y.}},
\bauthor{\bsnm{Miyazaki}, \binits{S.}}:
\bbtitle{Ni-Free Ti-Based Shape Memory Alloys}.
\bpublisher{Elsevier},
\blocation{Osford: UK}
(\byear{2018}).
\doiurl{10.1016/c2015-0-06056-7}
\end{bbook}
\endbibitem

\bibitem[\protect\citeauthoryear{Kong et~al.}{2022}]{KON-WAN-22}
\begin{barticle}
\bauthor{\bsnm{Kong}, \binits{L.}},
\bauthor{\bsnm{Wang}, \binits{B.}},
\bauthor{\bsnm{Sun}, \binits{S.}},
\bauthor{\bsnm{Hang}, \binits{X.}},
\bauthor{\bsnm{Meng}, \binits{X.}},
\bauthor{\bsnm{Zheng}, \binits{Y.}},
\bauthor{\bsnm{Gao}, \binits{Z.}}:
\batitle{Microstructure, superelasticity and elastocaloric behavior of {T}i-18{Z}r-11{N}b-3{S}n strain glass alloys by thermomechanical treatment}.
\bjtitle{Journal of Alloys and Compounds}
\bvolume{905},
\bfpage{164237}
(\byear{2022})
\doiurl{10.1016/j.jallcom.2022.164237}
\end{barticle}
\endbibitem

\bibitem[\protect\citeauthoryear{Seiner et~al.}{2023}]{KWINKS}
\begin{barticle}
\bauthor{\bsnm{Seiner}, \binits{H.}},
\bauthor{\bsnm{Sedl\'{a}k}, \binits{P.}},
\bauthor{\bsnm{Frost}, \binits{M.}},
\bauthor{\bsnm{\v{S}ittner}, \binits{P.}}:
\batitle{Kwinking as the plastic forming mechanism of b19' niti martensite}.
\bjtitle{International Journal of Plasticity}
\bvolume{168},
\bfpage{103697}
(\byear{2023})
\doiurl{10.1016/j.ijplas.2023.103697}
\end{barticle}
\endbibitem

\end{thebibliography}

\end{document}